\documentclass[10pt,aps,prc,floatfix,twocolumn,nofootinbib,superscriptaddress]{revtex4-2}
\usepackage{graphicx,amsmath,amssymb,bm}
\usepackage{amsfonts}

\usepackage[utf8]{inputenc}
\usepackage{verbatim}
\usepackage{float}
\usepackage[labelfont={small},font=small,subrefformat=parens,caption=false]{subfig}
\usepackage{cancel}
\usepackage{multirow}
\usepackage{array}
\usepackage{xparse}
\usepackage{xspace}

\usepackage{bigstrut}

\usepackage{physics}
\usepackage{color}
\usepackage{dsfont}

\usepackage{dcolumn}

\usepackage{xurl}
\usepackage[pdfencoding=auto, pdfpagelabels]{hyperref}

\usepackage[overload]{textcase}

\definecolor{linkcolor}{rgb}{0,0,0.40} 
\hypersetup{%
    pdfsubject=Paper,
    pdfkeywords={nuclear physics} {Bayesian} {chiral EFT} {three body force},
    unicode = true,
    breaklinks = true,
    colorlinks = true,
    linkcolor = linkcolor,
    citecolor = linkcolor,
    menucolor = linkcolor,
    urlcolor = linkcolor
}

\usepackage{cellspace}
\graphicspath{{./figures/}}

\makeatletter
\newcommand\newsubcommand[3]{\newcommand#1{#2\sc@sub{#3}}}
\def\sc@sub#1{\def\sc@thesub{#1}\@ifnextchar_{\sc@mergesubs}{_{\sc@thesub}}}
\def\sc@mergesubs_#1{_{\sc@thesub#1}}

\newcommand\newsupcommand[3]{\newcommand#1{#2\sc@sup{#3}}}
\def\sc@sup#1{\def\sc@thesup{#1}\@ifnextchar^{\sc@mergesups}{^{\sc@thesup}}}
\def\sc@mergesups^#1{^{\sc@thesup#1}}
\makeatother

\DeclareMathAlphabet{\mathbcal}{OMS}{cmsy}{b}{n}

\newcommand{\cbar}{\bar c}

\newcommand{\ordervec}{\vec}

\newcommand{\inputvec}{\mathbf}

\newsubcommand{\ckvec}{\ordervec{c}}{k}

\newsubcommand{\bkvec}{\ordervec{b}}{k}

\newsubcommand{\ckvecset}{\ordervec{\inputvec{c}}}{k}

\newsubcommand{\ckvecapprox}{\mathbf{c}'}{k}
\newsubcommand{\ckvecapproxset}{\mathbf{C}'}{k}

\newsubcommand{\bkvecapprox}{\mathbf{b}'}{k}
\newsubcommand{\bkvecset}{\mathbf{B}}{k}
\newsubcommand{\bkvecapproxset}{\mathbf{B}'}{k}

\newcommand{\genobs}{y}

\newsubcommand{\genobsvec}{\ordervec{\genobs}}{k}
\newsubcommand{\genobsvecset}{\ordervec{\inputvec{\genobs}}}{k}

\newsubcommand{\akvec}{\mathbf{a}}{k}

\newsubcommand{\akvecapprox}{\mathbf{a}'}{k}
\newsubcommand{\akvecset}{\mathbf{A}}{k}
\newsubcommand{\akvecapproxset}{\mathbf{A}'}{k}

{}  

\newcommand{\normal}{\mathcal{N}}

\def\diffd{\mathrm{d}}  

\DeclareDocumentCommand\differential{ o g d() }{ 
    \IfNoValueTF{#2}{
        \IfNoValueTF{#3}
            {\diffd\IfNoValueTF{#1}{}{^{#1}}}
            {\mathinner{\diffd\IfNoValueTF{#1}{}{^{#1}}\argopen(#3\argclose)}}
        }
        {\mathinner{\diffd\IfNoValueTF{#1}{}{^{#1}}#2} \IfNoValueTF{#3}{}{(#3)}}
    }

\newcommand{\pathd}{\mathcal{D}}  

\DeclareDocumentCommand\pathdifferential{ o g d() }{ 
    \IfNoValueTF{#2}{
        \IfNoValueTF{#3}
            {\pathd\IfNoValueTF{#1}{}{^{#1}}}
            {\mathinner{\pathd\IfNoValueTF{#1}{}{^{#1}}\argopen(#3\argclose)}}
        }
        {\mathinner{\pathd\IfNoValueTF{#1}{}{^{#1}}#2} \IfNoValueTF{#3}{}{(#3)}}
    }

\newcommand{\fdag}{f_{\dagger}}

\newcommand{\model}{\mathcal{M}}

\newcommand{\observations}{\mathbf{D}}

\begin{document}


\title{Interpolating between small- and large-$g$ expansions using Bayesian Model Mixing}

\author{A.~C. Semposki}
\email{as727414@ohio.edu}
\affiliation{Department of Physics and Astronomy and Institute of Nuclear and Particle Physics, Ohio University, Athens, OH 45701, USA}

\author{R.~J. Furnstahl}
\email{furnstahl.1@osu.edu}
\affiliation{Department of Physics, The Ohio State University, Columbus, OH 43210, USA}

\author{D.~R. Phillips}
\email{phillid1@ohio.edu}
\affiliation{Department of Physics and Astronomy and Institute of Nuclear and Particle Physics, Ohio University, Athens, OH 45701, USA}

\date{\today}

\begin{abstract}
Bayesian Model Mixing (BMM) is a statistical technique that can be used to combine models that are predictive in different input domains into a composite distribution that has improved predictive power over the entire input space. 
We explore the application of BMM to the mixing of two expansions of a function of a coupling constant $g$ that are valid at small and large values of $g$ respectively. This type of problem is quite common in nuclear physics, where physical properties are straightforwardly calculable in strong and weak interaction limits or at low and high densities or momentum transfers, but difficult to calculate in between.
Interpolation between these limits is often accomplished by a suitable interpolating function, e.g., Padé approximants, but it is then unclear how to quantify the uncertainty of the interpolant. 
We address this problem in the simple context of the partition function of zero-dimensional $\phi^4$ theory, for which the (asymptotic) expansion at small $g$ and the (convergent) expansion at large $g$ are both known. We consider three mixing methods: linear mixture BMM, localized bivariate BMM, and localized multivariate BMM with Gaussian processes. 
We find that employing a Gaussian process in the intermediate region between the two predictive models leads to the best results of the three methods. The methods and validation strategies we present here should be generalizable to other nuclear physics settings. 
\end{abstract}

\maketitle


\section{Motivation and goals} \label{sec:intro}

In complex systems, such as nuclei, there is typically more than one theoretical model that purports to describe a physical phenomenon. Among these models may be candidates that work well in disparate domains of the input space---energy, scattering angle, density, etc. In such a situation it is not sensible to select a single ``best'' model. How, instead, can we formulate a combined model that will be optimally predictive across the entire space?

A concrete realization of this situation is the combination of predictions from expansions about limiting cases, which provide theoretical control near those limits.
Nuclear physics examples include low- and high-density limits (e.g., quark-hadron phase transitions in dense nuclear matter~\cite{Glendenning:1992vb, Glendenning:2001pe, McLerran:2018hbz, Han:2019bub}); strong and weak interaction limits (e.g., expansions in positive and negative powers of the Fermi momentum times scattering length for cold atoms~\cite{Wellenhofer:2020ykf}); high- and low-temperature limits (e.g., Fermi-Dirac integrals~\cite{Johns:1996ht}); and four-momentum transfer limits (e.g., the generalized GDH sum rule~\cite{CLAS:2021apd}).
    
We propose using Bayesian statistical methods, and the technique of Bayesian model mixing (BMM) in particular, to interpolate between such expansions.
Bayesian methods are well suited to this problem since they enable the inclusion of prior information and provide principled uncertainty quantification (UQ). For a brief introduction to the basics of Bayesian statistics, see Appendix~\ref{bayesbackground}.

\begin{figure}[b]
    \centering
    \includegraphics[width=\columnwidth]{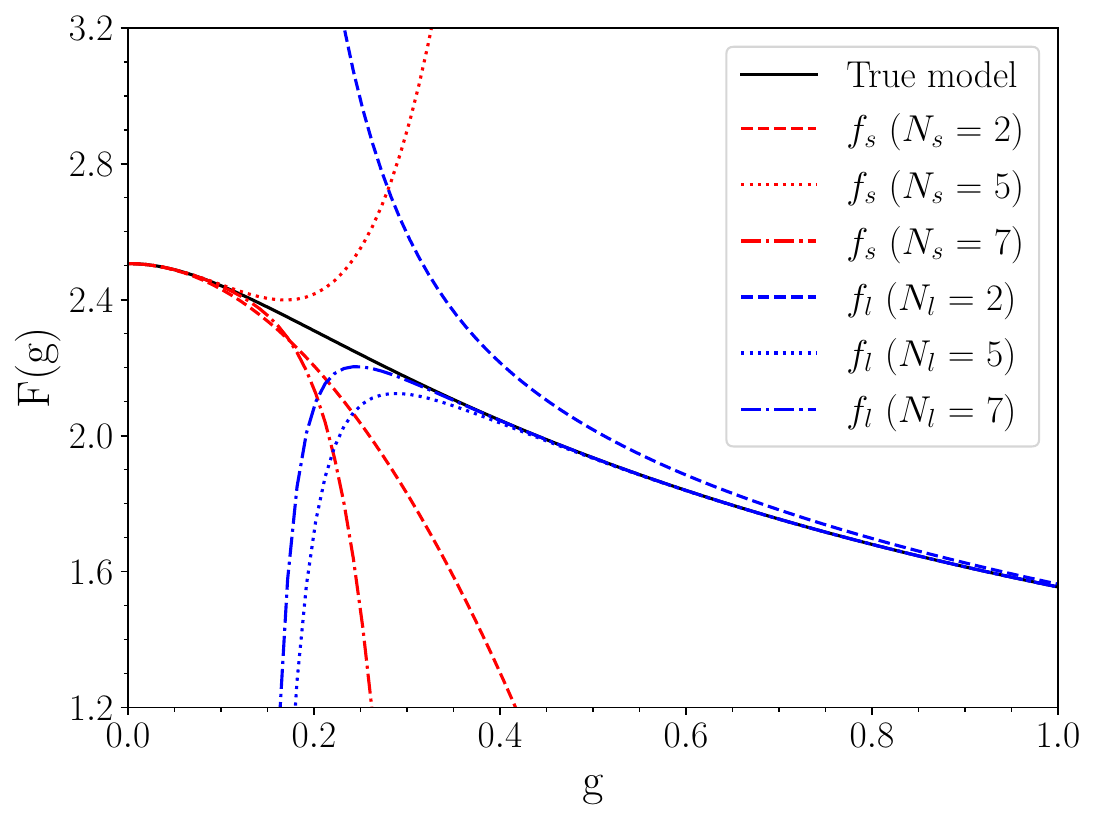}
    \caption{The partition function in Eq.~\eqref{eq:phi4}, denoted the true model, overlaid with two series expansions (red denoting the weak coupling expansion and blue the strong coupling expansion) truncated at various orders, here labelled $N_{s}$ and $N_{l}$ (see Eq.~\eqref{eq:expansions}).}
    \label{fig:multmodels}
\end{figure}

In this work, we explore BMM strategies for interpolant construction in the case of the partition function of zero-dimensional $\phi^{4}$ theory as a function of the coupling strength $g$ (see Eq.~\eqref{eq:phi4}).
These series expansions---in $g$ and $1/g$ respectively---provide excellent accuracy in their respective regions but, depending on the level of truncation, can exhibit a considerable gap for intermediate $g$ where they both diverge (see Fig.~\ref{fig:multmodels}). 

A variety of resummation methods have been used to extend the domain of convergence of the small-$g$ expansion of the zero-dimensional partition function of $\phi^4$ theory, although only a few have incorporated the constraints from the large-$g$ expansion; Ref.~\cite{Wellenhofer:2020ykf} discusses some of the possibilities. 
In Ref.~\cite{Honda:2014bza}, this zero-dimensional partition function was used to test the suitability of a generalized form of Pad\'e approximants as an interpolant between the small-$g$ and large-$g$ expansions. The resulting ``Fractional Powers of Rational Functions'' perform well in that case, but there is no obvious principle that guides the choice of the ``best'' rational power and, relatedly, without access to the true partition function, the accuracy of different interpolants is not accessible in the region of the gap. The same criticism applies to the constrained extrapolations with order-dependent mappings introduced in Ref.~\cite{Wellenhofer:2020ylh}. 

Here we will apply Bayesian methods to
 obtain an interpolant that agrees with the true model not only at both ends of the input space where the power series are reliable, but also in that gap. A key objective is that the interpolant should come with a  robust estimate of the uncertainty.
A hallmark of a Bayesian approach is the use of probability distributions for (almost) every component of the problem. 
Particularly useful here will be the 
posterior predictive distribution, or PPD, which is the distribution of possible values that a function can take at a point where we do not have data.
    We will label the $K$ models under consideration by $\model_k$, $(k=1,\ldots, K)$. The $k^{\rm th}$ model is specified by what it predicts for an observation $y_i$ 
    at a point $x_i$:
    \begin{equation} \label{eq:stand_model}
       \model_k : y_i = f_k(x_i) + \varepsilon_{i,k} .
     \end{equation}
     In our case the function $f_k$ 
     is a deterministic physics model depending on a single input $x_i$, or a Gaussian Process used to improve the interpolation between models.
     The physics model will usually also depend on parameters $\theta_k$, and those parameters would then have to be estimated by calibration to some observed data set  $\observations$, consisting of pairs $(x_1,y_1),\ldots,(x_n,y_n)$, but that is a complication we do not consider here\footnote{We note that the BAND~\cite{bandframework} software package \texttt{surmise} is capable of emulating and calibrating models and can be employed prior to using SAMBA to mix the models~\cite{surmise2021}.}. The error term, $\varepsilon_{i,k}$, represents all uncertainties (systematic, statistical, computational) which may also enter at the point $x_i$. Since $\varepsilon_{i,k}$ is a random variable, its probability distribution has to be specified in order to specify the statistical model for the relationship between $x_i$ and $y_i$.
      In our prototype, the $f_k$ are (different) expansions of the integral in Eq.~\eqref{eq:phi4}, with the $x_i$'s a set of coupling strengths $g$, and $\varepsilon_{i,k}$ modeled from the truncation error at $x_i$ for $f_k$. 
     
     The PPD for a new observation $(\tilde x, \tilde y)$ for model $k$ is denoted $p(\tilde y | \tilde x, \model_k)$.
     (Dependence on parameters $\theta_k$, if present, would be integrated out---``marginalized''---here.)
     Our goal is a combined PPD, which is a weighted sum of the PPD for each model:
     \begin{equation} \label{eq:PPD_weighting}
         p(\tilde y | \tilde x) = \sum_{k=1}^{K} \hat w_k p (\tilde y | \tilde x, \model_k).
     \end{equation}
     Conventional Bayesian Model Averaging (BMA)~\cite{Hoeting:1999abc} and Bayesian stacking~\cite{Yao:2018abc} are implementations of BMM for which the $\hat w_k$ are independent of $x_i$, but in general the weights $\hat w_k$ may depend on $\tilde x$~\cite{Yao:2021abc,BMM_local_10.2307/3088815}.
     Such dependence clearly has to be present for expansions that are valid in different input domains.
    
    The optimal ways to determine location-dependent weights in BMM for various applications is a topic of current research. The Bayesian Analysis of Nuclear Dynamics (BAND) project will produce a general software Framework for Bayesian analysis, with nuclear physics examples, and has a particular interest in the development and application of BMM methodologies.
    A general description of the BAND goals and BMM is given in the BAND Manifesto~\cite{Phillips:2020dmw}.
    
 The present work develops a theoretical laboratory (``sandbox'') that uses the example of the partition function in zero-dimensional $\phi^4$ theory to explore some possible approaches to BMM that produce reliable interpolants with principled UQ. We begin this enterprise in
Sec.~\ref{sec:formalism}, where we present in detail the prototype case we have chosen, and provide a concise description of Bayesian methods in the context of our problem. Following this,  in Sec.~\ref{sec:LMM}, we discuss BMM based on linear mixtures of individual models' PPDs  and present results of such an approach. In Sec.~\ref{sec:gaussians} we apply an alternative method, that we dub ``localized bivariate mixing'', which uses the series expansions and their variances to produce a precision-weighted combination of the small-$g$ and large-$g$ expansions. As an improvement upon this method, we introduce a Gaussian process as a third model, and present  results for ``localized multivariate mixing'' with that Gaussian process in Sec.~\ref{sec:gps}. A summary of lessons learned and the outlook for future applications of these techniques, as well as a short discussion of our publicly available SAndbox for Mixing using Bayesian Analysis (SAMBA) computational package, are contained in Sec.~\ref{sec:summary}. 

\section{Formalism} \label{sec:formalism}

\subsection{Test problem} \label{subsec:test_problem}

Our prototype true model is a zero-dimensional $\phi^{4}$ theory partition function~\cite{Honda:2014bza}
\begin{equation}
    \label{eq:phi4}
    F(g) = \int_{-\infty}^{\infty} dx~ e^{-\frac{x^{2}}{2} - g^{2} x^{4}} = \frac{e^{\frac{1}{32 g^{2}}}}{2 \sqrt{2}g} K_{\frac{1}{4}}\left(\frac{1}{32 g^{2}} \right),
\end{equation}
with $g$ the coupling constant of the theory, and $K_{\frac{1}{4}}$ the modified Bessel function of the second kind. This function can be expanded in either $g$ or $1/g$ to obtain:
\begin{equation} \label{eq:expansions}
F_{s}^{N_s}(g) = \sum_{k=0}^{N_{s}} s_{k} g^{k}, \qquad F_{l}^{N_{l}}(g) = \frac{1}{\sqrt{g}} \sum_{k=0}^{N_{l}} l_{k} g^{-k} ,
\end{equation}
with coefficients
\begin{equation}
s_{2k} = \frac{\sqrt{2} \Gamma{(2k + 1/2)}}{k!} (-4)^{k},~~~~~s_{2k + 1} = 0,
\end{equation}
and
\begin{equation}
l_{k} = \frac{\Gamma{\left(\frac{k}{2} + \frac{1}{4}\right)}}{2k!} \left(-\frac{1}{2}\right)^{k}.
\end{equation}
The weak coupling expansion, $F_{s}^{N_{s}}(g)$, is an asymptotic series, while the strong coupling expansion, $F_{l}^{N_{l}}(g)$, is a convergent series.  The former is a consequence of the factorial growth of the coefficients $s_{2k}$  at large $k$; as for the latter, the coefficients $l_k$ decrease factorially as $k \rightarrow \infty$. 

To apply Bayesian methods to the construction of an interpolant that is based on these small- and large-coupling expansions it is necessary to think of each expansion as a probability distribution. We follow Refs.~\cite{Cacciari:2011ze,Furnstahl:2015rha, Melendez:2019izc} and assign Gaussian probability distributions to the error made by truncating each series at a finite order. I.e., we define the PPD for the small-$g$ expansion at order $N_s$ as
\begin{equation}
{\cal M}_{N_s}: F(g) \sim {\cal N}(F_s^{N_s}(g),\sigma_{N_s}^2(g)),
\end{equation}
and that for the large-$g$ expansion at order $N_l$ as
 \begin{equation}
{\cal M}_{N_l}: F(g) \sim {\cal N}(F_l^{N_l}(g),\sigma_{N_l}^2(g)).
\end{equation}
Note that by assuming that the mode of the PPD is the value obtained at the truncation order we are assuming that the final result is equally likely to be above or below that truncated result.

\subsection{Truncation error model}

\label{sec:theoryerrormodel}

This leaves us with the task of defining the ``1-$\sigma$ error'', i.e., the standard deviation, associated with each PPD.  
The error in $F_s^{N_s}(g)$ is nominally of $O(g^{N_s+2})$ (note that the series contains only even powers of $g$) but the factorial behavior of the coefficients modifies the size of the error dramatically. We therefore build information on the asymptotics of the coefficient $s_{2N_s+2}$ into our model for that error. Analyzing the first few terms of the series (see Appendix~\ref{ap:truncationerrormodeldevelopment}) we might guess:
\begin{equation}
\sigma_{N_s}(g)=\left\{ \begin{array}{lc}
	\Gamma(N_s+3) g^{N_s + 2} \cbar, & \mbox{if $N_s$ is even;}\\
        \Gamma(N_s+2) g^{N_s+1} \cbar, & \mbox{if $N_s$ is odd.}
        \end{array} \right. 
        \label{eq:sigmauninformative}
\end{equation}
The standard deviation takes this alternating form because the small-$g$ series contains only even terms.
 $\cbar$ is a number of order one that is estimated by taking the (non-zero) terms in the expansion up to order $N_s$, rescaling them according to Eq.~(\ref{eq:sigmauninformative}) so that the large-$N_s$ behavior is accounted for, and computing their r.m.s. value. We deem Eq.~(\ref{eq:sigmauninformative}) an ``uninformative error model'' for large $N_{s}$. 
 
A closer look at the behavior of the series as $N_s$ gets bigger leads to a more informed statistical model for the uncertainty at order $N_s$:
\begin{equation}
\label{eq:informativesigmaNs}
\sigma_{N_s}(g)=\left\{ \begin{array}{lc}
	\Gamma(N_s/2+1) (4g)^{N_s + 2} \cbar, & \mbox{if $N_s$ is even;}\\
    \Gamma(N_s/2+1/2) (4g)^{N_s+1} \cbar, & \mbox{if $N_s$ is odd.}
    \end{array} \right. 
\end{equation}
$\cbar$ is again estimated by taking the non-zero terms in the expansion, this time rescaling them according to Eq.~(\ref{eq:informativesigmaNs}), and computing their r.m.s. value. Further discussion of how we arrived at these two forms for the behavior of higher-order terms can be found in Appendix~\ref{ap:truncationerrormodeldevelopment}.

Similarly the error in $F_l^{N_l}(g)$ is nominally of $O(g^{-N_l-3/2})$ (recall that $F(g)$ goes as $g^{-1/2}$ at large $g$) but we must first account for other  factors that dramatically affect the convergence of the series. We therefore take
\begin{equation}
    \sigma_{N_l}(g)=\frac{1}{\Gamma(N_l+2)} \frac{1}{g^{N_l+3/2}} \bar{d}
\end{equation}
as an uninformative model of the error at large $N_l$ and 
\begin{equation}
\sigma_{N_l}(g)=\left(\frac{1}{4g}\right)^{N_l + 3/2} \frac{1}{\Gamma(N_l/2+3/2)} \bar{d},
\label{eq:informativesigmaNl}
\end{equation}
as the informative model.
$\bar{d}$ is to be estimated using the same approach taken for $\bar{c}$.

It is important to note that all sets of coefficients discussed are alternating, so an error model with mean zero is appropriate. However, the error model only builds in some aspects of the series' behavior: it assumes that the higher-order coefficients are randomly distributed around zero, which does not account for their alternating-sign pattern. Some consequences of this choice will later be manifest in Fig.~\ref{fig:errorcomparison}.


\section{Linear mixture model} \label{sec:LMM}

\subsection{Formulation}

In some situations the true probability distribution is a linear mixture of two simpler probability distributions. For example, the height of adult males and the height of adult females\footnote{``Male'' and ``female'' here refer to the sex assigned at birth.} in the US are each approximately normally distributed~\cite{Ellenberg:2014ma}, but the two probability distributions have different means and variances, and so the distribution of adult heights in the US is (at the same level of approximation) a linear mixture of the two:
\begin{align}
& p({\rm height~of~US~adults})=\alpha p({\rm height~of~US~males})\nonumber\\
& \qquad \qquad \null + (1- \alpha)p ({\rm height~of~US~females}),
\end{align}
where $\alpha$ is the probability of drawing a male from the population. 

This kind of linear mixture of probabilities defines a conceptually simple form of BMM, cf.\ Eq.~(\ref{eq:PPD_weighting}). If the weights in Eq.~(\ref{eq:PPD_weighting}) are chosen to be independent of the location in the input space and equal to the Bayesian evidence (the denominator in Eq.~(\ref{eq:bayes})), what results is an example of BMA.

In this section we follow Coleman~\cite{Coleman} and make the weights in Eq.~(\ref{eq:PPD_weighting}) dependent on location in the input-space variable, $g$. Reference~\cite{Coleman} applied this technique to mix two different models of heavy-ion collisions developed by the JETSCAPE collaboration (although there the mixing protocol was also modified to account for correlations between observables across the input space). 

We apply it to large-$g$ and small-$g$ models, ${\cal M}_{N_l}$ and ${\cal M}_{N_s}$. The PPD computed in this section therefore takes the form
\begin{align}
    p(F(g)|\bm{\theta}) &\sim \alpha(g;\bm{\theta}) F_s^{N_s}(g) \nonumber\\
    & \quad\null +  (1-\alpha(g;\bm{\theta})) F_l^{N_l}(g).
    \label{eq:linear_mixture_model}
\end{align}
We take the variance of the two individual-model probability distributions to be the form of the informative error model (see Eqs.~\eqref{eq:informativesigmaNs} and \eqref{eq:informativesigmaNl}). The mixing function $\alpha(g;\bm{\theta})$ depends not just on the location in the input space, $g$, but also on parameters $\mathbf{\theta}$ defining that dependence on $g$. These parameters dictate how the mixing function ``switches'' from being 1 at $g=0$, where $F_s^{N_s}$ dominates, to being 0 at $g=1$, where $F_l^{N_l}$ dominates. 

\begin{figure}[h]
    \centering
    \includegraphics[width=\columnwidth]{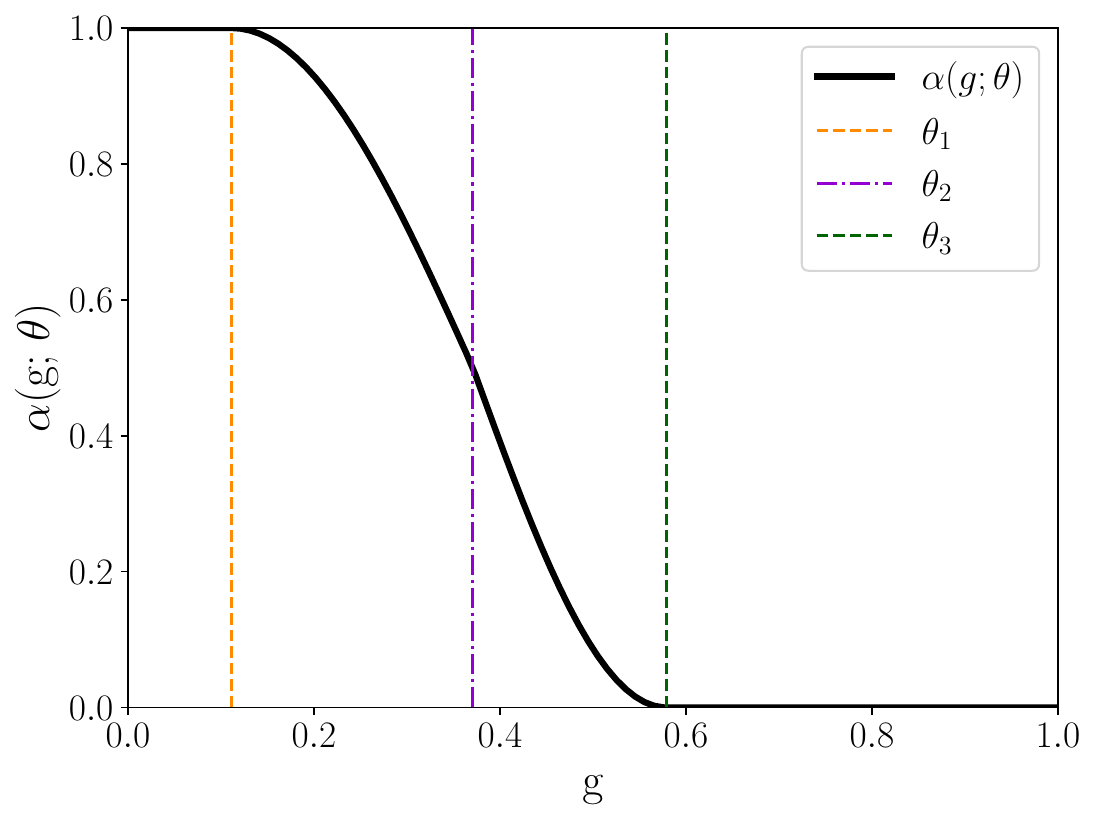}
    \caption{The piecewise cosine mixing function, evaluated at the MAP values of $\bm{\theta}$, shown as the solid black curve, for the case $N_{s}=N_{l}=2$. The dashed green, dashed orange, and dash-dotted purple lines represent the MAP values of the three parameters $\theta_1$, $\theta_2$, and $\theta_3$, which give the location of transition points in the function (see Eq.~\eqref{eq:cosine}).}
    \label{fig:pwcosine}
\end{figure}

We employ parametric mixing, which means we must specify the functional form of $\alpha$. In Ref.~\cite{Coleman} Coleman employed the well-known logistic function (which is also referred to as a sigmoid function, and has been used as a simple activation function in recent machine learning applications) and the CDF (cumulative distribution function) of the normal distribution. However, these do not meet the requirements of our application. We require $\alpha(g;\mathbf{\theta})$ become strictly 0 before the value of $g$ where the small-$g$ power series diverges. Likewise, $\alpha(g;\mathbf{\theta})$ must remain constant at 1 for all $g$ for which the large-$g$ expansion diverges. We therefore adopted a third mixing function, henceforth referred to as the cosine mixing function, which is constructed piecewise so that it switches the large-$g$ model on at a value $\theta_1$ that is ``late enough'', and switches the small-$g$ model off at a value $\theta_3$ that is ``early enough''. 
Figure~\ref{fig:pwcosine} provides a pictorial example of our function and its success in this constraint. The function is
\begin{widetext}
     \begin{equation}
         \label{eq:cosine}
         \alpha(g; \bm{\theta}) = 
         \begin{cases} 
          1, & g \leq \theta_{1}; \\
          \frac{1}{2}\left[1 + \cos(\frac{\pi}{2} \left(\frac{g-\theta_{1}}{\theta_{2} - \theta_{1}}\right))\right], & \theta_{1} < g \leq \theta_{2}; \\
          \frac{1}{2}\left[1 + \cos\left(\frac{\pi}{2} \left(1 + \frac{g - \theta_{2}}{\theta_{3} - \theta_{2}} \right) \right) \right],  & \theta_{2} < g \leq \theta_{3}; \\
          0, & g > \theta_{3}.
       \end{cases}
     \end{equation}
     \end{widetext}
The parameter $\theta_2$ specifies the location in input space, i.e., the value of $g$, at which each model has a weight of 0.5 in the final PPD.

The mixing function thus has three parameters $\bm{\theta}$ that need to be estimated from data. 
Bayes' theorem, Eq.~\eqref{eq:bayes}, tells us that the posterior for the parameters  is the product of the likelihood of that data and the prior distribution for the parameters. If we assume that we have $i=1,\ldots,n$ evenly spaced data points in the gap
and the error on these is $\sigma_{d_i}^2$ then we can use Bayes' theorem and the mixed-model definition (Eq.~\eqref{eq:linear_mixture_model}) to write
    \begin{align}
        \label{eq:lmmpost}
        p(\bm{\theta}|\mathbf{D}) &= p(\bm{\theta})\prod_{i=1}^{n} \left\{ \alpha(g_{i}; \bm{\theta}) ~\mathcal{N}(F^{N_s}_{s}(g_{i}), \sigma_{d_{i}}^{2} + \sigma_{N_s}^2) \right. \nonumber \\
         & \quad\null + \left. (1 - \alpha(g_{i}; \bm{\theta}))~ \mathcal{N}(F^{N_l}_{l}(g_{i}), \sigma_{d_{i}}^{2} + \sigma_{N_l}^2) \right\}.
    \end{align}
    Note that $\sigma_{N_s}^{2}$ and $\sigma_{N_l}^{2}$ correspond to the theory error model chosen (here the informative model). The prior distribution, $p(\bm{\theta})$, is defined as
    \begin{align}
        \label{eq:lmmprior}
        p(\bm{\theta}) &= \mathcal{U}(\theta_{1} \in [0,b])~\mathcal{N}(\theta_{1}; \mu_{1}, 0.05^2) \nonumber \\
        &\quad\null\times\mathcal{U}(\theta_{2} \in [\theta_{1},b])~\mathcal{N}(\theta_{2}; \mu_{2}, 0.05^2) \nonumber \\
        &\quad\null\times\mathcal{U}(\theta_{3} \in [\theta_{2},b])~\mathcal{N}(\theta_{3}; \mu_{3}, 0.05^2).
    \end{align}
This prior ensures that $\theta_1 < \theta_2 < \theta_3$. The value $b$ is a cutoff that can be selected to indicate where we believe the end of the mixing region between the models is (heretofore and henceforth deemed ``the gap"). This can be set to separate values in each uniform prior, depending on the user's knowledge of their system, or to the same value for all parameter priors. Note that this can indeed be part of the prior in this analysis, because we have access to information on where the gap between the expansions is based solely on the expansions themselves. Also, note that the values for the mean of the Gaussian priors ($\mu_1$, $\mu_2$, and $\mu_3$ in Eq.~\eqref{eq:lmmprior} above) can be assigned depending on the estimated location of the gap.

We determine (Eq.~\eqref{eq:lmmpost}), and hence the posteriors of the three parameters $\bm{\theta}$, by employing MCMC sampling, implemented via the Python package \texttt{emcee}~\cite{Foreman_Mackey:2013aa}. We assigned 20 walkers with 3000 steps each, for a total of 60,000 steps. We selected the first 200 steps of each walker as burn-in and removed them from consideration in the parameter chains. 

We can write the PPD (Eq.~\eqref{eq:linear_mixture_model}) as
\begin{align}
p(\tilde y(g)|\bm{\theta}, \mathbf{D}) &= \sum_{j=1}^{M} \alpha(g; \bm{\theta_{j}}) F^{N_s}_{s}(g) \nonumber \\
& \quad\null + (1 - \alpha(g; \bm{\theta_{j}})) F^{N_l}_{l}(g),
\label{eq:PPD_LMM}
\end{align}
where we sum over $M$ values of the $\mathbf{\theta}$ vector from trace results in our MCMC samples. In this way we are not just including a maximum a posteriori (MAP) value of our estimated parameters, but using all of the posterior distribution information we gained from our MCMC calculation to determine them.

\subsection{Results \& discussion}

\begin{figure}[h]
    \centering
    \includegraphics[width=\columnwidth]{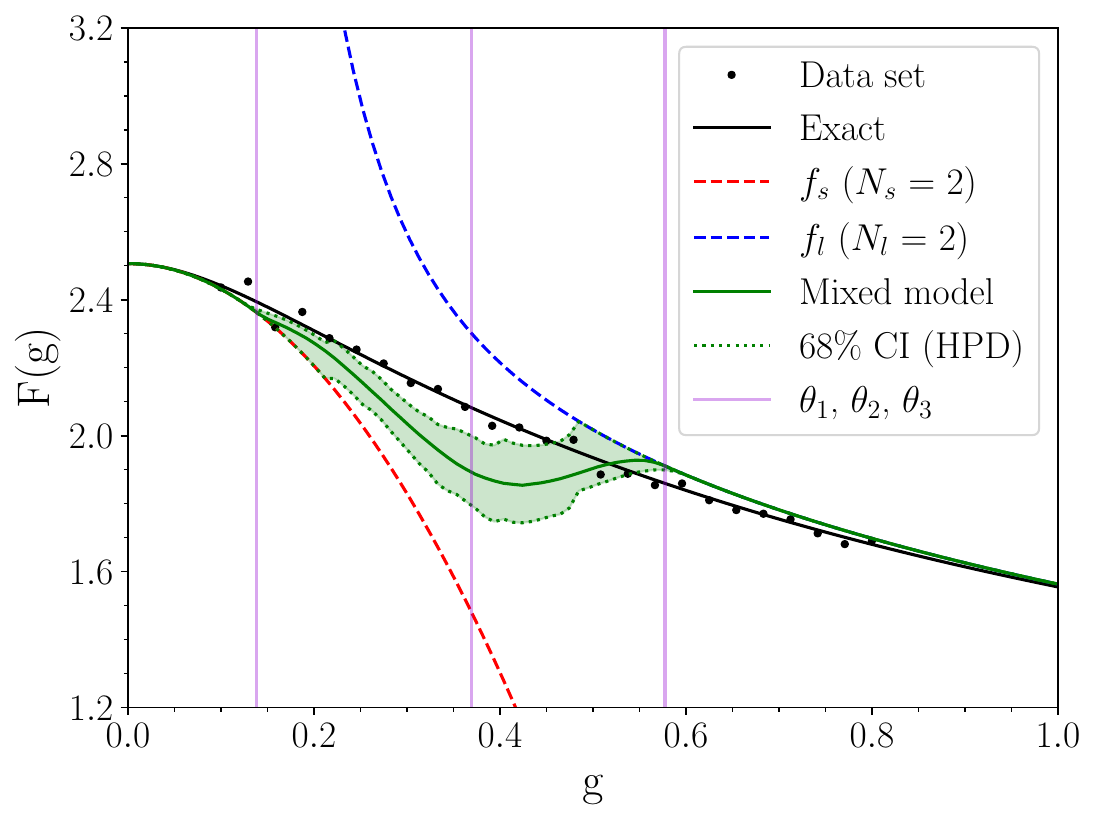}
    \caption{The PPD median and 68\% credibility interval for the linear mixture model (Eq.~\eqref{eq:PPD_LMM}) are shown in solid green and as a shaded green band, respectively, overlaid on the $N_{s}$ = 2 (dashed red curve) and $N_{l}$ = 2 (dashed blue curve) small-$g$ and large-$g$ expansions. The location in $g$ of the three mixture parameters of the piecewise cosine mixture function, $\theta_1$, $\theta_2$, and $\theta_3$, are indicated by the solid light purple lines. The true model curve is shown in solid black as a reference, with the black dots around it comprising the randomly distributed data set used in this analysis. This data has a 1\% error from the true curve added to it.}
    \label{fig:pwclmm22}
\end{figure}

\begin{figure}[h]
    \centering
    \includegraphics[width=\columnwidth]{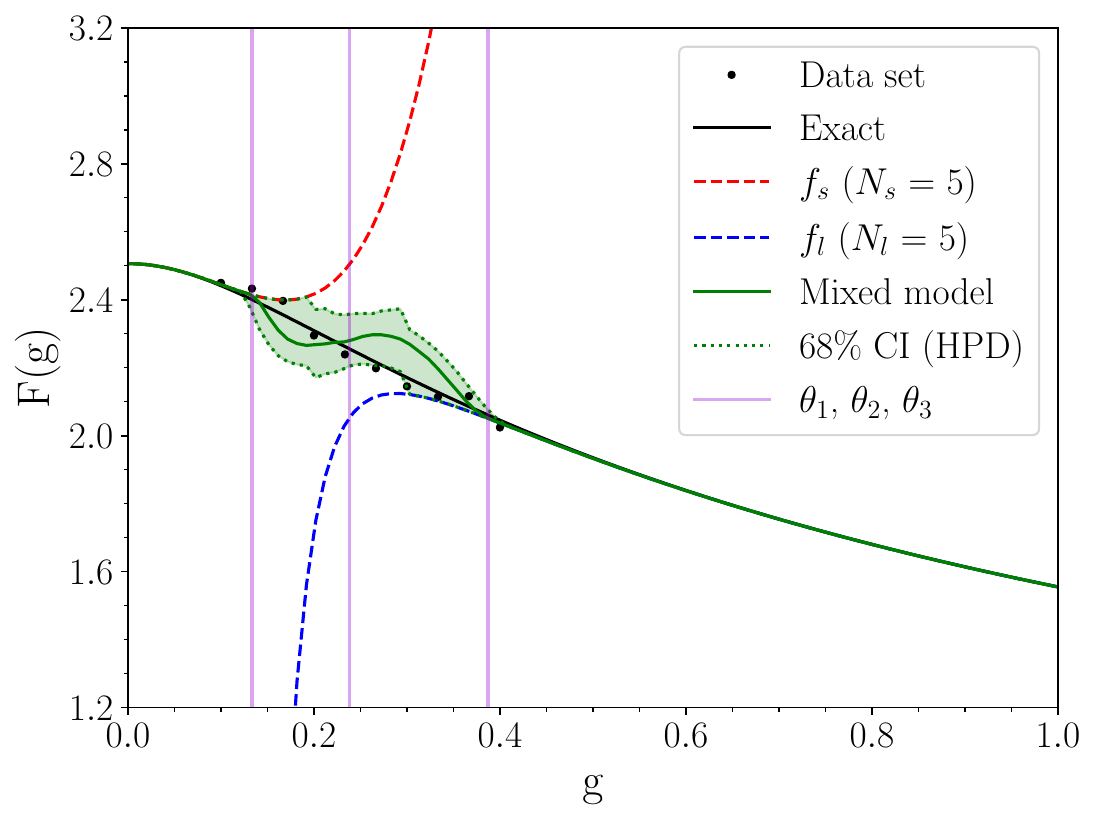}
    \caption{The PPD median and 68\% credibility interval, with all of the curves as for Fig.~\ref{fig:pwclmm22}, but for the case $N_s = N_l = 5$.}
    \label{fig:pwclmm55}
\end{figure}

Before discussing the results for the linear mixture model we point out that the need for data in the gap between these two expansions to calibrate the parameters of the mixing function could be a concern for applications to physical problems where obtaining this data would be costly, difficult, or both. 

Even if data is available---as we have assumed it is---two obvious pitfalls of the linear mixture model in our test case can be seen in Figs.~\ref{fig:pwclmm22} ($N_s=N_l=2$) and~\ref{fig:pwclmm55} ($N_s=N_l=5$). The PPD mean connects to the two series expansions across the gap, but in both cases the discontinuities in the first derivatives of the mixed model at the left and right edge of the gap would not easily be explained by natural occurrence, unless one crossed a phase transition or other new physics entered at a specific value of $g$. 

Second, even though the theory error of the small- and large-$g$ expansions (in the form of the informative error model used in this analysis) is accounted for in our analysis, we see that in Fig.~\ref{fig:pwclmm22} the 68\% credibility interval does not include the true model 68\% of the time, indicating that this method is not accurately quantifying uncertainty in this toy problem. We did find cases where this linear mixture model produces a PPD with good empirical coverage properties, i.e., does not suffer from this defect, but we were only able to achieve that for $N_{s}, N_{l} = 5$ (as in Fig.~\ref{fig:pwclmm55}) as well as for some values that are much higher than would be attainable in a physics application.

There is a separate, but perhaps related, issue in choosing which mixture function we use by hand. If we are not choosing one that really represents the transition from one model's domain to the next, we are building in an implicit bias to the mixed model result, as the model will always adopt features of the mixing function chosen. In the next section, we construct another method that addresses this problem directly by eliminating the need for a mixing function to cross the gap. 


\section{Localized bivariate Bayesian model mixing} \label{sec:gaussians}

\subsection{Formalism} \label{sec:bivformalism}

One way to formulate a mixed model that incorporates the model uncertainty of the models being mixed 
 is to use the simple method of combining Gaussian distributed random variables~\cite{Phillips:2020dmw}. This can be written for any number of models $K$. Each model is weighted by its \emph{precision}, the inverse of the variance:
\begin{equation} \label{eq:precisionweighting}
   \fdag = \frac{1}{Z_P}\sum_{k=1}^{K} \frac{1}{v_k}f_k,
   \qquad Z_P \equiv \sum_{k=1}^{K}\frac{1}{v_k} .
\end{equation}
From the rule for a linear combination of independent normally distributed random variables $X_k \sim \normal(\mu_k,v_{k})$,
\begin{equation} \label{eq:combinedgaussgen}
    \sum_k a_k X_k \sim \normal\Bigl(\sum_k a_k \mu_k, \sum_k a_k^2 v_{k}\Bigr) ,
\end{equation}
the distribution of $\fdag$ is
\begin{equation}
    \label{eq:fdagger}
    \fdag \sim \normal\bigl(Z_P^{-1}\sum_k \frac{1}{v_k}f_k, Z_P^{-1}\bigr) .
\end{equation}
The variance $v_{k}$ is determined by the error models previously discussed in Sec.~\ref{sec:theoryerrormodel}, thereby incorporating the theory uncertainty into this mixing method. This section focuses on the bivariate case of Eq.~\eqref{eq:fdagger}, where model 1 and model 2 are $F_{s}^{N_{s}}(g)$ and $F_{l}^{N_{l}}(g)$, respectively. We employ the so-called ``informative" error model 
and several values of $N_{s}$ and $N_{l}$ (for more examples of $N_{s}$ and $N_{l}$, see the supplemental material for this work \ref{ap:supplemental}). 

\subsection{Results \& discussion}

\begin{figure}[thb]
    \centering
    \includegraphics[width=\columnwidth]{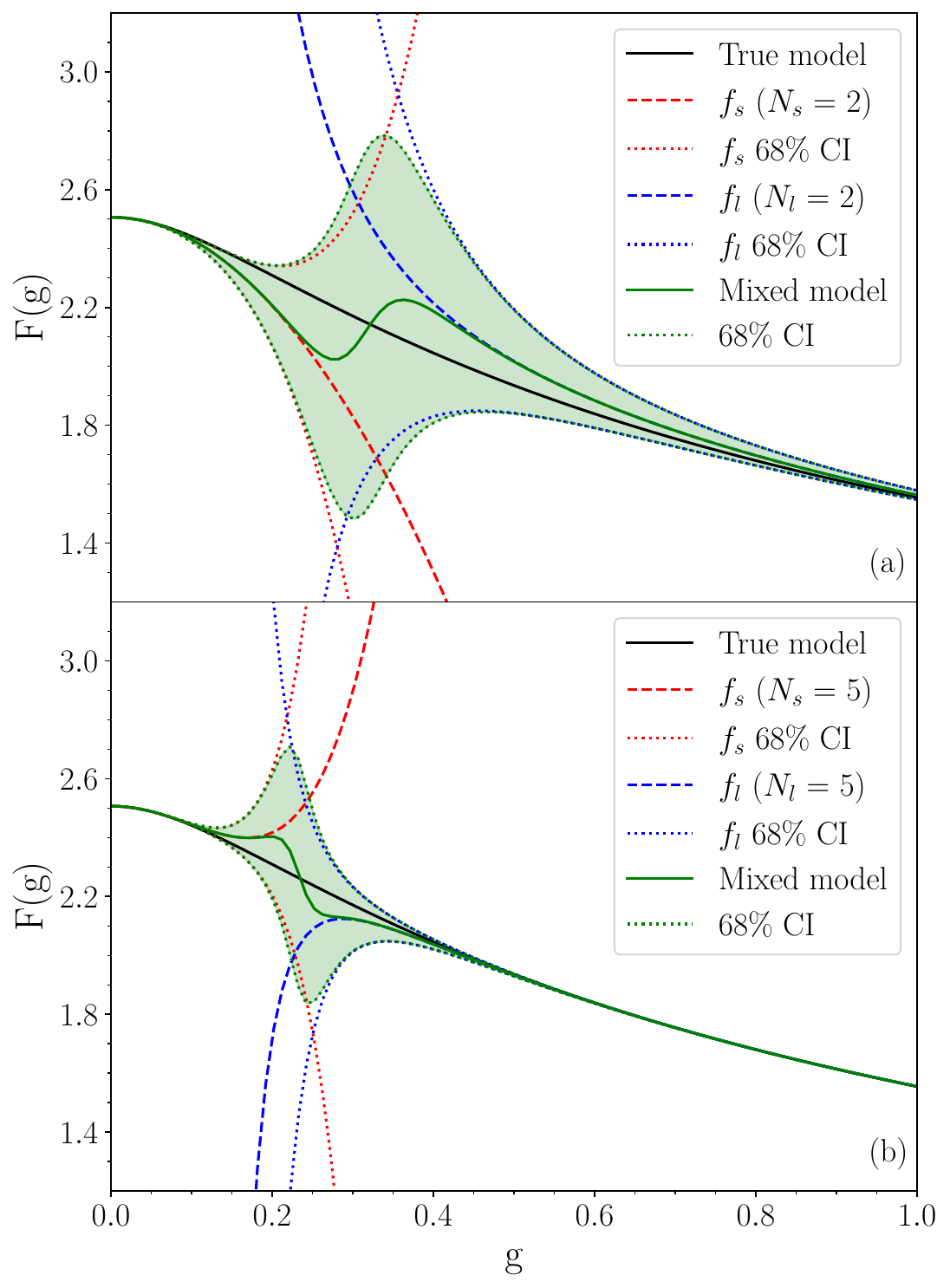}
    \caption{The results of computing the bivariate model mixing are shown for (a) $N_{s} = 2$, $N_{l} = 2$, and (b) $N_{s} = 5$, $N_{l} = 5$, using the informative error model. The result from computing $f_{\dagger}$ (Eq.~(\ref{eq:combinedgaussgen})) is shown in green, with the 68\% credibility interval shown as the green shaded region.}
    \label{fig:bivgauss_vertical}
\end{figure}

\begin{figure*}[tbh]
    \centering
    \includegraphics[width=\textwidth]{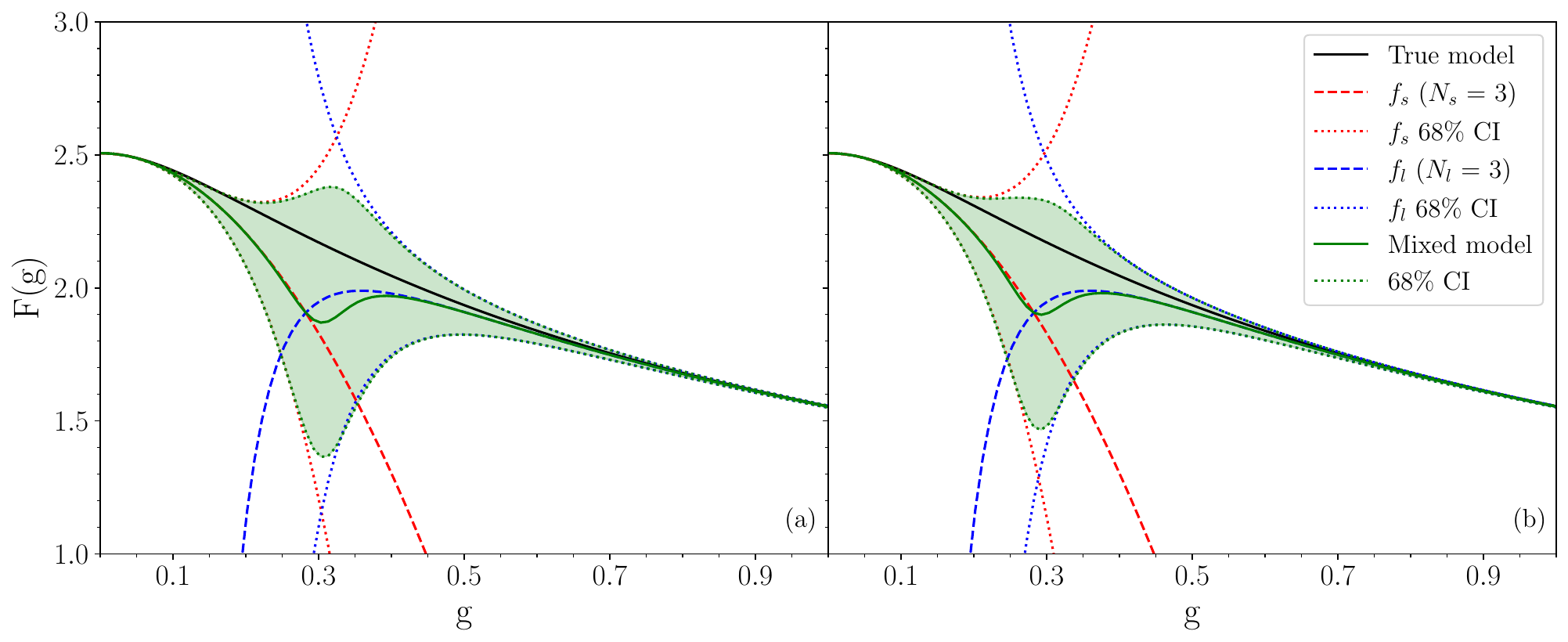}
    \caption{A comparison between the uninformative error model (left panel) and the informative error model (right panel), shown side by side applied to the case $N_{s}=3$, $N_{l}=3$. The greater size of the uncertainty in the gap when the uninformative error model is used is evident in the left panel.}
    \label{fig:errorcomparison}
\end{figure*}

The results of this method for two different cases of $N_{s}$, $N_{l}$ are presented in Fig.~\ref{fig:bivgauss_vertical}. The bivariate mixed model completes the journey across the gap for both the lower orders (panel (a)) and the higher orders (panel (b)) in a continuous fashion, as desired. In both cases the true curve is also encompassed by the 68\% interval of the PPD obtained through bivariate model mixing. 

However, there are two negative aspects of the result of Fig.~\ref{fig:bivgauss_vertical} which are quite striking. First, the shape of the mean of the PPD, is not in line with our intuition for this problem. Second, the 68\% interval is, if anything, too wide. Considering that $F$ only varies between 2.5 and 1.8 between $g=0$ and $g=0.6$ an error bar of order $0.3$, as is obtained even for $N_s=N_l=5$, seems rather conservative. 

In fact, as was discussed in Sec.~\ref{sec:bivformalism}, the right variances to use in Eq.~\eqref{eq:combinedgaussgen} to mix the two power series are nontrivial to determine. This adds to the challenges of this  technique. There is a significant effect of the form of the variance on the results of the bivariate mixed model---the credibility interval changes by a noticeable amount depending on whether we employ the so-called uninformative or the informative formulae (see Fig.~\ref{fig:errorcomparison} and Sec.~\ref{sec:theoryerrormodel}). Hence, there is a substantial amount of uncertainty---at least in the 68\% interval of the interpolant's PPD---from not knowing the correct form of the truncation error. This affects the final result obtained from this two-model mixing in a way that cannot be ignored. 

In fact, these ``failures'' all stem from the same root cause: the use of only the small-$g$ and large-$g$ power series of $F$ to form the interpolant. If this is truly all the information one has then this is all one can say~\cite{wittgenstein-1922}. Given the errors on the small- and large-$g$ expansions, behavior encompassed by the green band is possible. And nothing we have put into our bivariate model-mixing formulation precludes a sinusodial variation of the type seen in both panels of Fig.~\ref{fig:bivgauss_vertical}.

But such curvature of the mixed model mean is not a feature that is supported by any prior knowledge of our particular system. The intuitive expectation would be to see a smooth, linear path from one model to the other, and that is not being shown in our result from this simple two-model method.
In the next section we incorporate this expectation by including in our mixing a nonparametric form for the function $F$ in the gap between the two expansions.


\section{Bridging the gap: localized multivariate Bayesian model mixing with Gaussian processes} \label{sec:gps}

\subsection{Formalism}\label{sec:multiformalism}

To improve upon Sec.~\ref{sec:gaussians}, we introduce a third model to interpolate across the gap in $g$ in the form of a Gaussian process. A Gaussian process (GP) is fundamentally stochastic---a collection of random variables, any subset of which will possess a multivariate Gaussian distribution~\cite{Melendez:2019izc}. We can describe a GP with its mean function and positive semi-definite covariance function, or \textit{kernel}. The  kernel should reflect the  physical and mathematical characteristics of our system. In this work, we focused on two of the many possible kernel forms: the RBF (Radial Basis Function) kernel, and the Mat\'ern 3/2 kernel. The former is described by
\begin{equation}
    \label{eq:rbf}
    k(x,x') = \exp\left( -\frac{(x-x')^{2}}{2l^{2}} \right),
\end{equation}
while the latter is given by
\begin{align}
    \label{eq:matern}
    k(x, x') = \frac{1}{\Gamma(\nu)2^{\nu - 1}} &\left( \frac{\sqrt{2\nu}}{l}(x-x') \right)^{\nu} \nonumber \\
    & \null\times K_{\nu} \left( \frac{\sqrt{2\nu}}{l}(x-x') \right),
\end{align}
where $\Gamma$ is the gamma function, and $K_{\nu}$ is the modified Bessel function of the second kind. We employed the package \texttt{scikit learn} to compute our GP within the wrapper of SAMBA~\cite{scikit-learn}. General draws from both of these kernels can be seen in Sec.~1.7.5 of the \texttt{scikit learn} documentation (see~\cite{scikit-doc}). We chose the Mat\'ern 3/2 kernel specifically, hence $\nu=3/2$ in Eq.~\eqref{eq:matern}. 

Employing an emulator typically involves three steps: setting up a kernel, choosing and training the emulator on a data set, and predicting at new points using the trained GP. Our toy model is trained using four points from training arrays generated across the input space. The four points chosen are determined by: \begin{itemize}
    \item Generating an array of training points from both expansions.
    \item Selecting two fixed points from the small-$g$ expansion training array and the last fixed point in the large-$g$ expansion training array.
    \item Determining the fourth and final training point from the large-$g$ expansion training array using what we will call a training method.
\end{itemize}
We also include the truncation error on each point of the training set. This is done as symmetric point-to-point uncertainty, fixed using the informative error model.

In our case, we applied
three different training methods in the final step. Their results can be found in Table~\ref{tab:training_methods}. The first (referred to as method 1) fixes the fourth training point at $g$ = 0.6 for every value of $N_{s}$ and $N_{l}$ investigated. The second (method 2) allows the training point to shift given the values of $F_{s}(g)$ and $F_{l}(g)$, locating the final training point 
where the value of $F_s(g)$ becomes much larger than the range of output values in this problem. The third (method 3) employs the theory error on each training array point. This method fixes the fourth training point at the location in the data set where the value of the theory error at that point changes by more than 5\% of $F_l(g)$ at the next point. 

With the GP in hand we then use Eq.~\eqref{eq:fdagger} to mix it with $F_s^{N_s}(g)$ and $F_l^{N_l}(g)$ to obtain our mixed model. Results of this procedure are discussed in Sec.~\ref{sec:gpresults}.

\subsection{Diagnostics: Mahalanobis distance} \label{sec:diagnostics}

\begin{table*}[t]
    \centering
    \begin{ruledtabular}
    \begin{tabular}{c|c|c|c|c|c|c|c}
    & & \multicolumn{3}{c|}{$\textrm{D}_{\textrm{MD}}^{2}$ (Mat\'ern 3/2)} & \multicolumn{3}{c}{$\textrm{D}_{\textrm{MD}}^{2}$ (RBF)} \\
      & & \multicolumn{3}{c|}{Training Method} & \multicolumn{3}{c}{Training Method} \\
     $N_{s}$ & $N_{l}$  &  1 &  2 & 3 & 1 & 2 & 3 \\ 
     \hline
     2 & 2 & 1.6 & 1.4 & 1.6 & 60 & 70 & 60 \\
     2 & 4 & 1.1 & 0.6 & 5.0 & 140 & 170 & 1500 \\
     3 & 3 & 0.5 & 0.3 & 3.0 & 65 & 110 & 7 \\
     4 & 4 & 1.8 & 4.3 & 5.6 & 66 & 29000 & 600 \\
     5 & 5 & 2.7 & 1.2 & 1.2 & 350 & 15 & 15 \\
     8 & 7 & 1.9 & 0.4 & 0.8 & 4200 & 210000 & 14 \\
     5 & 10 & 5.1 & 3.4 & 6.0 & 1100 & 140 & 2100 \\
     \hline
     \multicolumn{2}{c|}{Mean} & 2.1 & 1.7 & 3.3 & 850 & 34000 & 610 \\
    \end{tabular}
    \end{ruledtabular}
    \caption{Results of the squared Mahalanobis distance for the Mat\'ern 3/2 kernel and RBF kernel for all three training methods discussed. The mean of each training method per kernel is also calculated. From this data, it is evident that the Mat\'ern 3/2 kernel is the most suitable for our problem.}
    \label{tab:training_methods}
\end{table*}

We employ the (squared) Mahalanobis distance to check which kernel and training method are optimal for this toy model. 
The Mahalanobis distance includes correlated errors and maps the differences of the GP and the true solution across the input space to a single value. It is given by 
\begin{equation}
    \label{eq:mahalanobis}
    D^{2}_{MD} = (\bm{y} - \bm{m})^{T}\textit{K}^{-1}(\bm{y} - \bm{m}) , 
\end{equation}
where $\bm{y}$ is a vector of solutions (either of a draw from the GP being used or the true solution) and $\bm{m}$, $\textit{K}$ are the mean and covariance matrix from the GP, as previously noted~\cite{Melendez:2019izc}.

To determine whether or not the calculated squared Mahalanobis distance is reasonable, we compare it to a reference distribution. To construct the reference distribution we take $\bm{y}$ to be numerous draws from a multivariate normal distribution constructed from the GP mean and covariance matrix. This reference distribution should converge to a $\chi^{2}$-distribution with the degrees of freedom matching the number of points used to calculate $\mathrm{D}_{\mathrm{MD}}^{2}$~\cite{BastosDiagnosticsGaussianProcess2009}. We used samples from the emulator to construct the reference distribution and compared to a $\chi^{2}$ with the appropriate degrees of freedom. This verifies that the reference distribution is indeed the $\chi^2$ distribution, so hereafter we compare our value of $\mathrm{D}_{\mathrm{MD}}^{2}$ to the expected value given the $\chi^{2}$. 

Now we take $\mathbf{y}$ to be the vector representing the true solution at a series of points in $g$, as we want to compare our model to the true result for this toy problem. We pick three points in $g$ to calculate the Mahalanobis distance, and ensure that they are sufficiently spaced that there are not too many points within one lengthscale of the GP. We obtain the results  for $\mathrm{D}_{\mathrm{MD}}^{2}$ for each training method and kernel given in Table~\ref{tab:training_methods}. Seven pairs of truncation orders $N_{s}, N_{l}$ were chosen to test each training method and kernel previously discussed. A suitable Mahalanobis distance is one that is close to the number of degrees of freedom of the $\chi^{2}$-distribution we are comparing to. Examining the RBF kernel results, it is obvious that this kernel is not the correct choice for this problem. The computed $\mathrm{D}_{\mathrm{MD}}^{2}$ is orders of magnitude too large in many cases there, regardless of training method. Because of this, we rule out using the RBF kernel entirely. The enormous squared Mahalanobis distance values given by this kernel seem to be due to its smoothness---the RBF kernel is so smooth that it breaks down and leads to $K$ in Eq.~\eqref{eq:mahalanobis} becoming asymptotically singular. Inverting this matrix then leads to these inflated $\mathrm{D}_{\mathrm{MD}}^{2}$ numbers. This can be seen by invoking the singular value decomposition (SVD). This shows a drop in the magnitude of the eigenvalues by several orders, progressing from the first to the last. Often the final eigenvalue is on the order of $10^{-7}$ or smaller, because the RBF covariance matrix is near degenerate~\cite{Ababou:1994}. 

Examining the training method results under the Mat\'ern 3/2 kernel, each of the calculated squared Mahalanobis distances is within the variance of the $\chi^{2}(N=3)$ distribution. This implies that the method used does not have an overly strong correlation with the result of the interpolation such as to be devastating if we chose, say, method 1 over method 2. In all of our final results we employed method 2, as it yields the smallest Mahalanobis distance of the three methods here.

\subsection{Results \& discussion} \label{sec:gpresults}

\begin{figure}[h]
    \centering
    \includegraphics[width=\columnwidth]{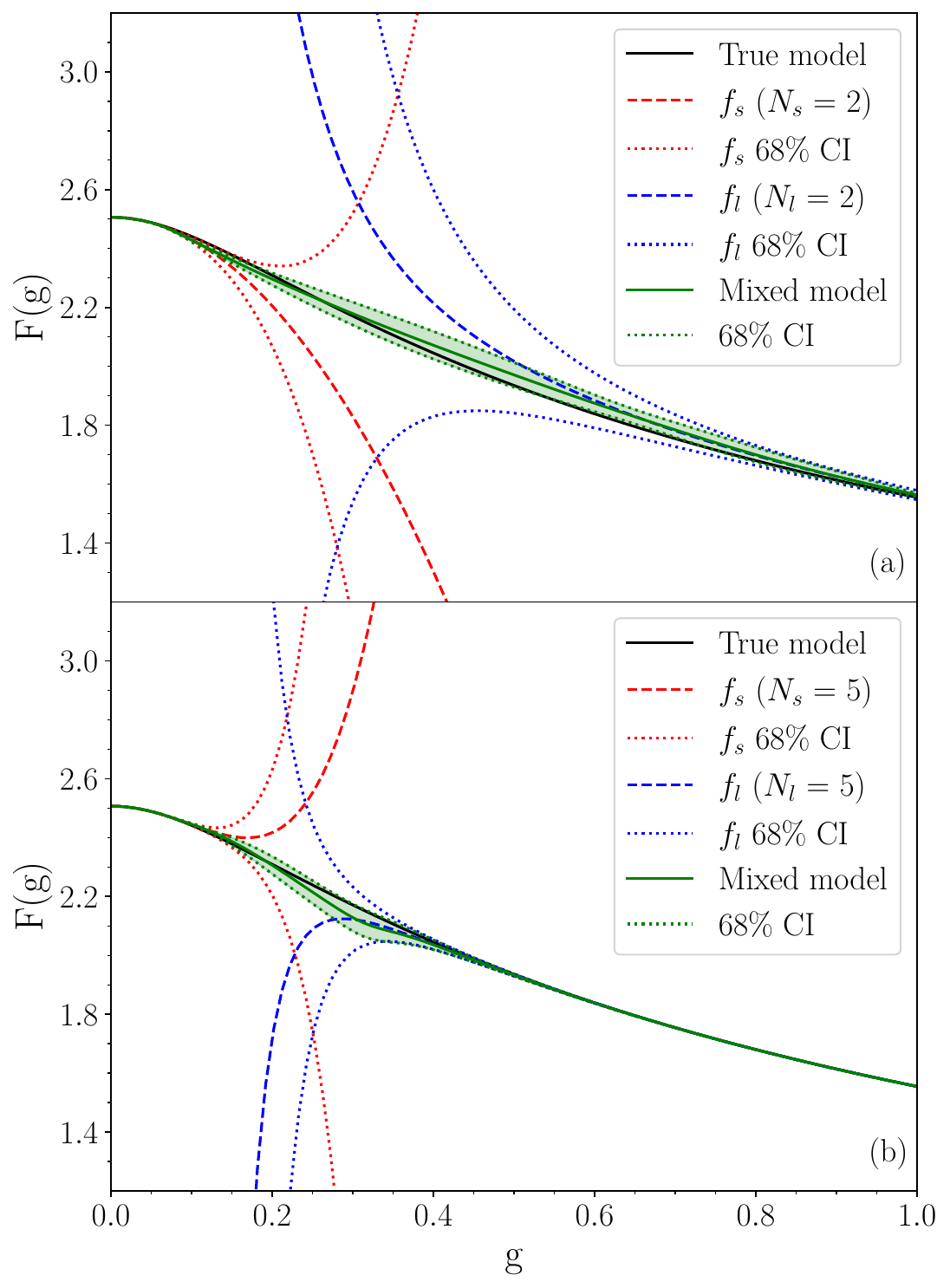}
    \caption{The results of employing the Mat\'ern 3/2 kernel in a GP to cross the gap between the two series expansions, truncated at (a) $N_{s}$ = 2, $N_{l}$ = 2, and (b) $N_{s}$ = 5, $N_{l}$ = 5. Compare this result to Fig.~\ref{fig:bivgauss_vertical}, which did not include the GP.}
    \label{fig:multigp_vertical}
\end{figure}

In Fig.~\ref{fig:multigp_vertical}, the mixed models have been plotted for $N_{s}, N_{l} = 2$ and $N_{s}, N_{l} = 5$ (panels (a) and (b), respectively). These truncation orders are the same as those in Fig.~\ref{fig:bivgauss_vertical}, chosen for ease of direct comparison between the two figures. Adding the GP as an interpolant is shown to greatly lessen the curvature of the mixed model across the gap, as well as suppress the uncertainty in that region to obtain a more precise estimate of the true curve. The improvement in UQ is so significant that the error model chosen (discussed in Secs.~\ref{sec:theoryerrormodel} and \ref{sec:gaussians}) no longer has a sizable effect on the results of the mixed model. This is excellent news in regards to generalising this method---if the final mixed curve does not display a lot of sensitivity to the high-order behavior of the coefficients in the theory error model the interpolants that result will be much more robust.

A complementary verification of our success in estimating the uncertainty of the mixed model is evident in Table~\ref{tab:gap_widths}, where the gap length (over $g$) is given for each truncation order pair, and the area in the uncertainty band is calculated via a simple implementation of Simpson's rule. The reduction of the gap area as the truncation order increases and the width decreases supports our GP result as a good interpolant for this mixed model, as the size of the uncertainty band should decrease the more terms (information) one adds to the mixed model.

\begin{figure*}
    \centering
    \includegraphics[width=\textwidth]{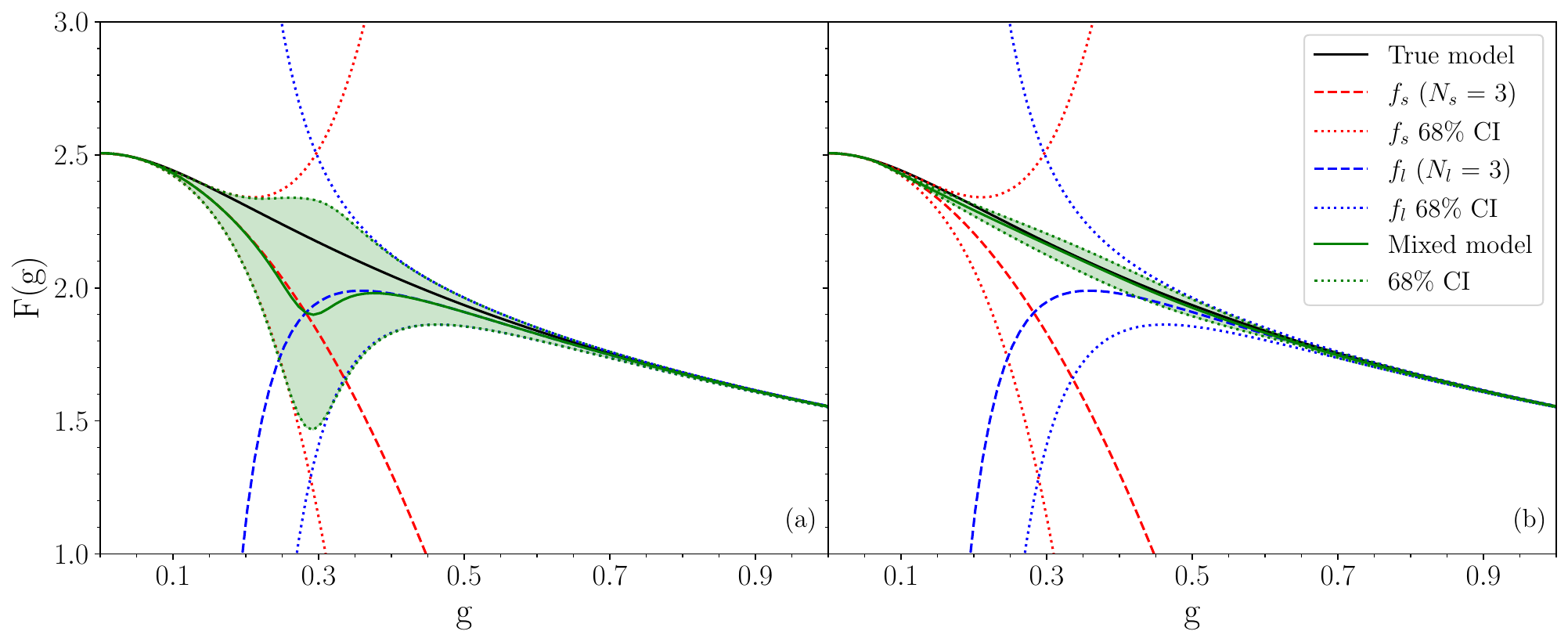}
    \caption{A comparison of the results of the method from Sec.~\ref{sec:gaussians} and that of Sec.~\ref{sec:gps}, applied to the $N_{s}$ = 3 and $N_{l}$ = 3 cases. The reduction in the credibility interval is evident in the latter method, as well as the lessening of the curvature in the gap between the expansions.}
    \label{fig:comparison33}
\end{figure*}

A direct comparison of the bivariate results to those of the multivariate model mixing using a GP for $N_{s}, N_{l}$ = 3 can be examined in Fig.~\ref{fig:comparison33}. In this case, the mixed model including the GP (panel (b)) is a very close approximation to the true curve, a stark contrast to the mixed model in panel (a), where the mixed curve follows the individual models much more closely than the true curve. The effect of the two expansions both approaching $-\infty$ is nearly nonexistent when the GP is employed for this case, showing that this method has greatly improved upon Sec.~\ref{sec:gaussians} in multiple ways. Both the reduction of the overly conservative error bar in panel (a) and the improvement of the mean curve are promising results. These further our argument that this method, suitably generalized, will be a great asset to anyone wishing to model mix across an input space such as this.

\begin{table}
    \centering
    \begin{ruledtabular}
    \begin{tabular}{c|c|c|c}
    $N_{s}$ & $N_{l}$ & Gap length ($g$) & Gap area \\
    \hline
    2 & 2 & 0.92 & 0.05 \\
    2 & 4 & 0.60 & 0.02 \\
    3 & 3 & 0.92 & 0.04 \\
    4 & 4 & 0.26 & 0.03 \\
    5 & 5 & 0.21 & 0.02 \\
    8 & 7 & 0.13 & 0.01 \\
    5 & 10 & 0.11 & 0.003 \\
    \end{tabular}
    \end{ruledtabular}
    \caption{Calculated areas of each uncertainty band per truncation order in $N_{s}$ and $N_{l}$. As expected, as both $N_s$ and $N_l$ increase, the uncertainty reduces.}
    \label{tab:gap_widths}
\end{table}

\subsection{Weights}
\label{sec:weights}

\begin{figure}[h]
    \centering
    \includegraphics[width=\columnwidth]{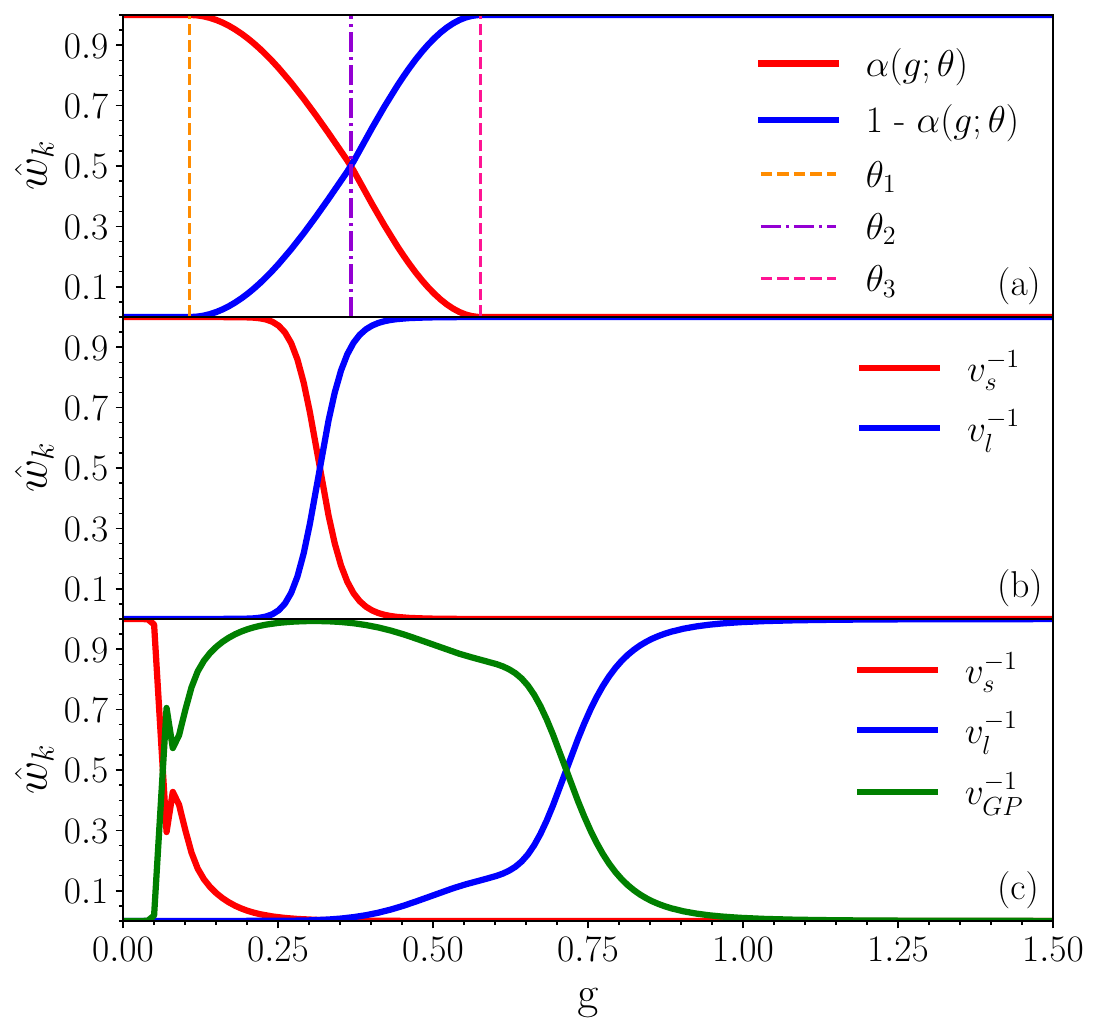}
    \caption{The normalized weights of each model ($N_{s}, N_{l} = 2$, shown in typical red and blue) computed across the input space $g$, for each method (panel (a): linear mixture model, panel (b): bivariate BMM, panel (c): trivariate BMM including the GP). Note the green in panel (c) is the weights curve for the GP interpolant.}
    \label{fig:weights22}
\end{figure}

\begin{figure}[h]
    \centering
    \includegraphics[width=\columnwidth]{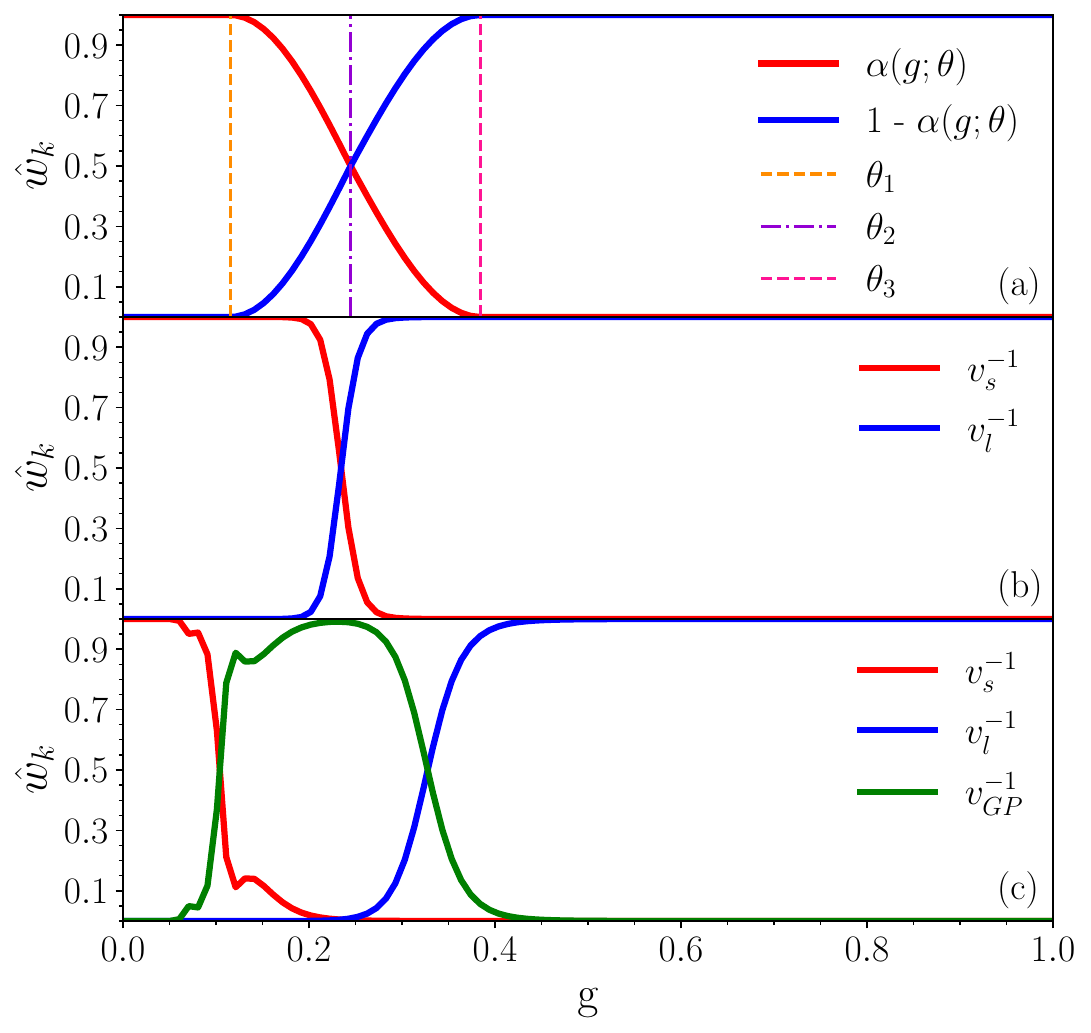}
    \caption{The normalized weights of each model as a function of the input space $g$, as in Fig.~\ref{fig:weights22}, though now for $N_{s}, N_{l} = 5$. Note the decrease in the region in which the GP dominates in this figure, in contrast to the case in Fig. \ref{fig:weights22}.}
    \label{fig:weights55}
\end{figure}

As previously discussed, within BMM it is possible to construct weights that are dependent upon the location in the input space considered: for our toy problem this means the weights are a function of $g$. We explicitly plot them for each BMM method used for the case of $N_{s} = N_{l} = 2$ (Fig.~\ref{fig:weights22}) and $N_{s} = N_{l} = 5$ (Fig.~\ref{fig:weights55}). These weights are plotted as $\alpha$ and ($1 - \alpha$) functions in the linear mixture model (panel (a) in both figures), and normalized precisions in the case of the bivariate BMM and trivariate BMM (panels (b) and (c) in both figures). 

The panel (a)s in these two figures both look like we would expect, given our knowledge of $\alpha(g;\bm{\theta})$ from Sec.~\ref{sec:LMM}: the red curve dominates until some crossing point where the blue takes over as the dominant weight, indicating that the most influential model on the mixed model turns slowly from the $F_s^{N_s}(g)$ (weighted by the red curve) to $F_l^{N_l}(g)$ (weighted by the blue curve). The same can be seen in panel (b) in Figs.~\ref{fig:weights22} and \ref{fig:weights55} where the two curves also progress in the same fashion as in panel (a), showing the crossover from $F_s^{N_s}(g)$ dominance at low $g$ to $F_l^{N_l}(g)$ dominance at large $g$. But, the crossover in the bivariate case is much sharper than that in the linear mixture model case. 

The panel (c)s highlight the substantial influence of the GP in the trivariate mixed model result across the gap---in the center of the gap for both Figs.~\ref{fig:weights22} and \ref{fig:weights55}, the GP reaches a weight of nearly 1.0, indicating it is exerting the most control over the result of BMM there. The decline of the GP curve is rather slow on the large-$g$ side compared to the sharp rise on the low-$g$ side of the input space, which could be related to the gap slowly closing on the large-$g$ end in comparison to its somewhat quicker opening on the small-$g$ end (see Fig.~\ref{fig:multigp_vertical}, panel (a) where this is evident). 
The location in the input space of the crossover from GP to $F_l^{N_l}(g)$ for the trivariate mixed model compared to when $F_l^{N_l}(g)$ takes over when we mix only two models varies with the orders considered.
In the $N_s=N_l=5$ case it is similar to the linear mixture model but for the $N_s=N_l=2$ case the GP dominates well beyond where the gap closes for two models.

\subsection{Comparison with the FPR method} \label{sec:fpr}
    
In Ref.~\cite{Honda:2014bza}  a method involving Fractional Powers of Rational Functions (FPR method) was used to interpolate between these two limits. The FPR method postulates a form
\begin{equation}
\label{eq:FPR}
F(g)=g^a \left(\frac{1 + c_1 g + c_2 g^2 + \ldots + c_p g^p}{1 + d_1 g + d_2 g^2 + \ldots + d_q g^q}\right)^\alpha,
\end{equation}
that is designed to mimic a Pad\'e approximant, but be more flexible since it allows an arbitrary leading power at small $g$ and quite a lot of flexibility as regards the choice of the fractional power $\alpha$ to which the Pad\'e will be raised.
If we have $N_s$ ($N_{l}$) coefficients from the small (large)-$g$ expansion then we need~\cite{Honda:2014bza}:
\begin{align}
p &=\frac{1}{2}\left(N_l+N_s+1-\frac{(a-b)}{\alpha}\right), \nonumber \\
q &=\frac{1}{2}\left(N_l+N_s+1+\frac{(a-b)}{\alpha}\right).
\end{align}
The $p$ and $q$ coefficients $c_k$ and $d_k$ are then adjusted to satisfy the constraints imposed by the need to reproduce the first $N_s$ terms of the small-$g$ expansion of $F$ and the first $N_l$ terms of the large-$g$ expansion of $F$. As is clear from Eq.~(\ref{eq:FPR}), the coefficient $a$ serves to ensure the small-$g$ behavior is correctly reproduced. In our specific toy problem, we possess $a = 0$ and $b = -1/2$. 

\begin{figure}[h]
    \centering
    \includegraphics[width=\columnwidth]{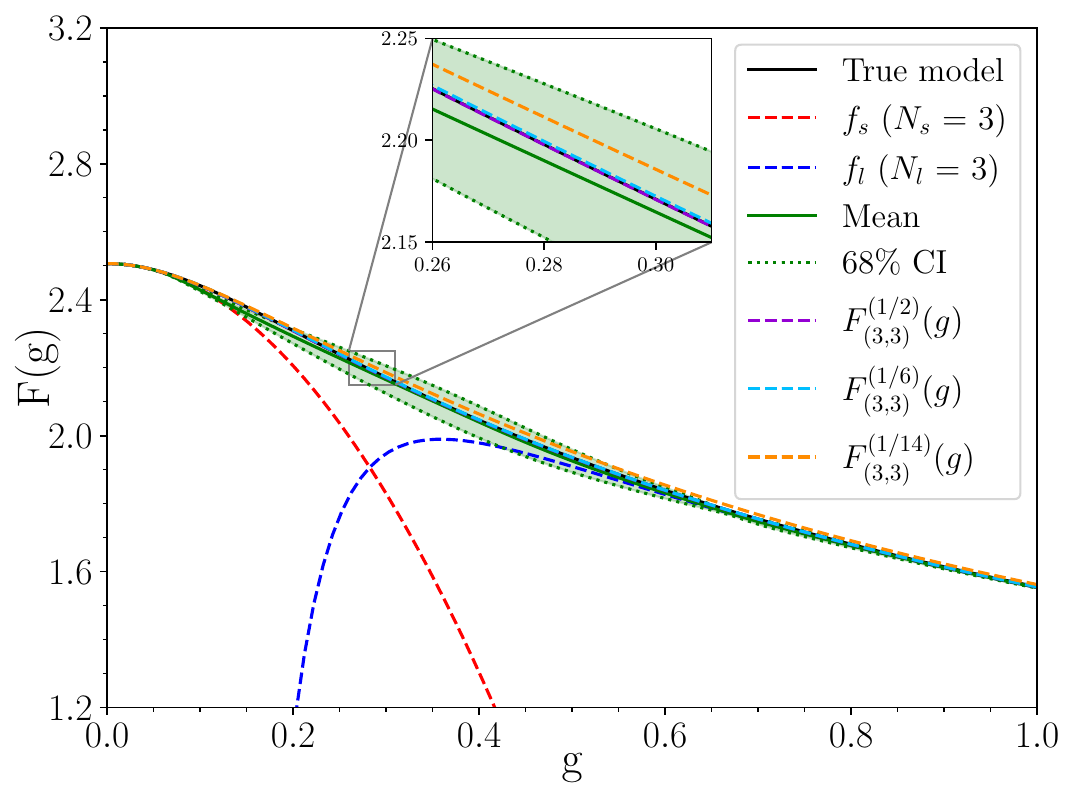}
    \caption{Overlay of the FPR method curves for different values of $\alpha$ in the $N_{s}$, $N_{l}$ = 3 case with the mixed model results using the GP. \emph{Inset}: A close up of the central mixing region, where it is easier to note the deviation of the FPR results from the true solution as the value of $\alpha$ decreases.}
    \label{fig:fprplot}
\end{figure}

In Fig.~\ref{fig:fprplot} we compare the results of Honda's FPR code~\cite{Honda:2014bza} with our multivariate mixed model for $N_{s}, N_{l} = 3$. Honda chooses three different $\alpha$ values, and as $\alpha$ decreases the accuracy of the FPR method decreases, as shown in the inset. Our GP mixed model mean is not as close to the true result as the FPR results for $\alpha$ = 1/2 and 1/6. Nevertheless, the FPR method does not include any UQ, which is arguably the most important development of this work, making our method 
a compelling option
for researchers wishing to include principled UQ in their results. Furthermore, while the FPR interpolation method applies only to series expansions, our approach can be adapted to many other situations.


\section{Summary and outlook} \label{sec:summary}

In this work we have explored how to apply Bayesian model mixing to a simple test problem, interpolating expansions in the coupling constant $g$ of a zero-dimensional $\phi^4$ theory partition function (see Sec.~\ref{subsec:test_problem}).
Rather than choose the single ``best'' model, as in Bayesian model selection, or an optimal global combination of predictions, as in BMA, we seek to vary the combination of models as a function of input location.
The goal is to optimize the predictive power across the full input domain and quantify the prediction's uncertainty.
Introducing BMM with a transition region is more flexible than treating this as a model selection problem. 
(Model selection can be considered a special case of BMM, for which the weights of each model are either zero or one.)

The basic output of BMM for our example and for more general cases is a combined PPD, see Eq.~\eqref{eq:PPD_weighting}, which is a probability distribution for the output at a new input point based on an appropriate weighted sum of individual PPDs.
To be clear, the prediction is not simply the most probable outcome, although the distribution may be summarized by MAP values and the variance (or covariances in general).  
The challenge is to how to best determine these input-dependent weights.
  
A general feature of this task is the need to incorporate error estimates from theory or learned from data. 
In our toy model,
we exploit analytic solutions for the full integral and for small $g$ and large $g$ expansions.
Analytic knowledge of these expansions enables credible error estimates; in practice we simulate different levels of credibility with ``uninformative'' and ``informative'' error models, see Fig.~\ref{fig:errorcomparison} for an example.
By changing the order of each expansion, we can adjust the gap between their individual regions of highest validity as a function of $g$.
Thus this problem serves as a prototype for combining predictive models 
with different patterns of coverage across the full input domain of interest.

We have explored a sequence of BMM strategies.
In Sec.~\ref{sec:LMM} we considered linear mixtures of PPDs, following Coleman~\cite{Coleman}, using the informative error model and parametric mixing with a switching model (see Fig.~\ref{fig:pwcosine}).
We identified pitfalls with this approach
when it is necessary to ``mind the gap''; in particular, when the gap is large, this approach does not yield a statistically consistent result for the intermediate region. The linear mixture model did produce a PPD with good empirical coverage properties for high-order expansions. Perhaps unsurprisingly, it is easier to bridge the gap when the gap is small.

While reaching such high orders is unlikely to be realistic for physics applications, linear mixing may be an appropriate procedure when the domains of applicability overlap significantly. Many discussions about what to do in the region between the two expansions become moot if the expansions share a region in which both are convergent. 
Competing EFT expansions provides an example, such as those for pionless and chiral effective field theory (EFT) in the few-nucleon problem.

In Sec.~\ref{sec:gaussians} we considered localized bivariate mixing, with each model weighted by its precision (inverse variance) to incorporate the theory uncertainty. 
Here we found good but rather conservative empirical coverage and a non-intuitive mean prediction for our model problem: where we expected a more-or-less straight joining of the limiting expansions the mean curves were significantly curved.
These are not, however, failures because they correctly reflect the limited prior information about the behavior of the function in the gap. Since
we used only the small-$g$ and large-$g$ expansions to inform the mixture, sinusoidal behavior of the mean is not ruled out, and is seen in Fig.~\ref{fig:bivgauss_vertical}. The small- and large-$g$ expansions themselves do not  preclude a rapid change in the region where neither is valid.

Taking care not to over-constrain intermediate behavior may be relevant for the  application to bridging the nuclear matter equation of state between saturation density and QCD ``asymptopia".
At the low density end, chiral EFT provides an expansion with theoretical truncation errors, while at the high density end one can anchor the result with a QCD calculation, as in Ref.~\cite{Huth:2020ozf}. 
Because the question of an intermediate phase transition is open, we would want to minimize prior constraints. 

Ambiguities as regards the behavior in the intermediate regime will disappear in applications where competing expansions overlap---as already pointed out for linear mixing.
Even if expansions do not overlap we may have prior information that the observable being studied is smooth in the intermediate region. 
In Sec.~\ref{sec:gps} we introduced a model to bridge the gap in the toy model with prior expectations of smoothness, implemented as localized multivariate mixing using a GP.
Not unexpectedly, the details of the GP, such as the choice of the kernel, matter (Mat\'ern 3/2 is much favored over RBF), but the use of a Mahalanobis diagnostic to check on the kernel and training method shows reasonable insensitivity to the choice of the latter.
The final results are very good.
If we did not seek an uncertainty estimate, the FPR results~\cite{Honda:2014bza} might be favored for highest precision in this particular application.
But, for the more likely extensions to nuclear physics problems, our multivariate model mixing is a compelling approach.

When using this approach to construct the mixed model we did not take into account correlations between the truncation error in each of the power series across different values of $g$. Such correlations do not affect the mixing at a particular value of $g$. Extending the multivariate model mixing approach employed here to obtain the covariance matrix of the mixed model that emerges from combining $K$ models, each of which has a non-diagonal covariance matrix, is an interesting topic for future investigation.   

We expect that 
our explorations could be extended to many different phenomena in nuclear physics.
For example, another candidate for BMM is to bridge the gap between the situations in nuclear data where $R$-matrix theory dominates and where Hauser-Feshbach statistical theory is the principal choice. 
This and other applications will require extending the BMM considered here to multidimensional inputs, as has been done in the case of Bayesian model averaging (e.g., \cite{Kejzlar:2020vla}).
The SAMBA computational package will enable new practitioners to learn the basics and to explore strategies in a simple setting. Not only can users try the toy model detailed in this work, but they will be able to include their own data and try these model mixing methods for themselves in the next release. 
This will include adoption of other BMM strategies, such as using Bayesian Additive Regression Trees (BART)~\cite{chipman2010bart,bart_doi:10.1146/annurev-statistics-031219-041110}.


\begin{acknowledgments}
We thank \"Ozge S\"urer, Jordan Melendez, Moses Chan, Daniel Odell, John Yannotty, Matt Pratola, Tom Santner, and Pablo Giuliani for helpful discussion on many sections of this work, and suggestions with regards to the structure of SAMBA. We thank Frederi Viens for supplying names for the methods of     Secs.~\ref{sec:gaussians} and \ref{sec:gps} and Kostas Kravvaris for suggesting an interesting application of this approach.
This work was supported by the National Science Foundation Awards No. PHY-1913069 and PHY-2209442 (R.J.F.), CSSI program Award No. OAC-2004601 (A.C.S., R.J.F., D.R.P.), and the U.S. Department of Energy under Contract No. DE-FG02-93ER40756 (D.R.P.)
\end{acknowledgments}


\appendix
\section{Bayesian Background}
\label{bayesbackground}

We provide a simple description of the mechanics of Bayesian statistics (see~\cite{Phillips:2020dmw} and \cite{Neufcourt:2018syo} for a more comprehensive exposition). Bayes' theorem relates two different conditional probability distributions: 
    \begin{equation}
        \label{eq:bayes}
        p\mathrm{(\bm{\theta}|\mathbf{D})} = \frac{p\mathrm{(\mathbf{D}|\bm{\theta})} p\mathrm{(\bm{\theta})}}{p\mathrm{(\mathbf{D})}},
    \end{equation}
    i.e., it relates the likelihood determined from data $\mathbf{D}$ and the prior information on parameters $\bm{\theta}$ to the posterior of $\bm{\theta}$. The evidence term in the denominator of Eq.~(\ref{eq:bayes}) serves as the normalization of this relation and can be dropped if one is only interested in working within a single model.

The prior probability distribution in Eq.~(\ref{eq:bayes}), $p\mathrm(\bm{\theta})$, is a specific representation of any prior knowledge we may have about the parameters we are seeking. It is imperative that we include any physical information known about the system in this quantity when applying Bayesian methods to real physics problems: this is the way we can include physical constraints in the statistical framework.
    
A commonly used method to determine the posterior density distributions of parameters $\bm{\theta}$ is Markov Chain Monte Carlo (MCMC) sampling. This technique often utilizes a Metropolis-Hastings algorithm to map out the posterior of each parameter given the likelihood and prior PDFs. We employed this technique in Sec.~\ref{sec:LMM}.

\section{Truncation error model development aka how to be a good guesser aka how to be a poor guesser at first and then get better}

\label{ap:truncationerrormodeldevelopment}

We take two models $f_1$ and $f_2$, with variances $v_1$ and $v_2$, and form the pdf for the mixed model, $f_\dagger$, according to
\begin{equation}
    f_\dagger \sim {\cal N}\left(\frac{v_1 f_2 + v_2 f_1}{v_1 + v_2},\frac{v_1 v_2}{v_1 + v_2}\right),
\end{equation}
as detailed in Sec. 3.4 of Ref. \cite{Phillips:2020dmw}.
In our implementation the variances are defined by the error model associated with omitted terms. 

\textit{Uninformative error model.} For $F_s(g)$, the expansion in positive powers of $g$, we have coefficients that behave as:
\begin{align}
\frac{s_n}{n!}=\{&2.5066, -3.7599, 5.4833, -6.0316, \nonumber \\
&5.2507, -3.7688, 2.2984,-1.2178,\nonumber\\ 
&0.5702, -0.2391, 0.09081, -0.03150, \nonumber \\
&0.01006, -0.002975, 0.0008193; \nonumber \\ & n=2, 4, 6\ldots, 30\},
\end{align}
with even more rapid decrease thereafter. All odd powers of $g$ have vanishing coefficients. Taking the first-omitted term approximation to the part of the expansion omitted at order $2n$ we can adopt
\begin{equation}
    \sigma_{N_s} = \bar{c} (N_s+2)! g^{N_s+2}
\end{equation}
for $N_s$ even and
\begin{equation}
        \sigma_{N_s} = \bar{c} (N_s+1)! g^{N_s+1}
\end{equation}
if $N_s$ is odd.
Here the coefficient $\bar{c}$ is estimated by taking the root-mean-square value of coefficients up to the order of truncation $N_s$. This will have a tendency to overestimate the next coefficient, but it provides a conservative error bar.

For $F_l(g)$, the expansion in negative powers of $g$, we have coefficients that behave as:
\begin{align}
l_m*m!=\{&1.813, -0.3064, 0.1133, -0.05744, \nonumber \\ 
& 0.03541, -0.02513, 0.01992, -0.01728,  \nonumber\\
&0.01618, -0.01620, 0.01719,-0.01923, \nonumber \\
&0.02257, -0.02765, 0.03526; \nonumber \\
&m=0,\ldots,14\},           
\end{align}
with eventual rapid increase. Here we once again take the first-omitted term approximation to $v_2$ to estimate the error due to truncation at order $N_l$ as:
\begin{equation}
    \sigma_{N_l}=\bar{d} \frac{1}{(N_l+1)!} \frac{1}{g^{N_l+3/2}}.
\end{equation}
Here, $\bar{d}$ is estimated by taking the root-mean-square value of coefficients from order $2$ to the $m$th order~\footnote{The first two coefficients shift $\bar{d}$ to a value much higher than any of the succeeding coefficients would indicate. This, in turn, widens the interval on the large-g expansion much more than we would wish to observe, so in this model we neglect them.}. There might be some tendency to underestimate the next coefficient if we are unlucky as to where we stop. But, up to order $m=14$, this seems to provide a reasonable estimate of the size of the next term.

\textit{Informative error model.} Let us suppose we are better guessers, as regards the asymptotic behavior of the coefficients. Then we might form 
\begin{align}
\frac{s_{2n}}{(n-1)! 4^{2n}}=\{&-0.469993, 0.514055, -0.530119, \nonumber \\
&0.538402, -0.543449, 0.546846, \nonumber\\
&-0.549287, 0.551126, -0.552562, \nonumber \\ &0.553713;\nonumber \\
&n=2,4,\ldots,20\}, 
\end{align}
which is remarkably stable out to $n=60$. Similarly, we might look at:
\begin{align}
l_m*\Gamma(m/2+1)*4^m=\{&1.8128, -1.086, 0.9064,\nonumber \\
&-0.8145, 0.7553,-0.7127, \nonumber\\
  &0.6798,-0.6533, 0.6312, \nonumber \\
  &-0.6125, 0.5962, -0.5818; \nonumber \\ &n=0,\ldots,12\} 
\end{align}
which has only a very slow decrease, reaching $0.3968$ by $n=50$. Based on these behaviors we formulate error model 2, as defined by Eqs.~(\ref{eq:informativesigmaNs}) and (\ref{eq:informativesigmaNl}).

\section{Supplemental material \label{ap:supplemental}}

\subsection{Linear Mixture Model}

\begin{figure*}[htbp]
  \subfloat[Truncation orders $N_{s}=5$ and $N_{l}=10$.]{
	\begin{minipage}[c][1\width]{
	   0.45\textwidth}
	   \centering
	   \includegraphics[width=1.1\textwidth]{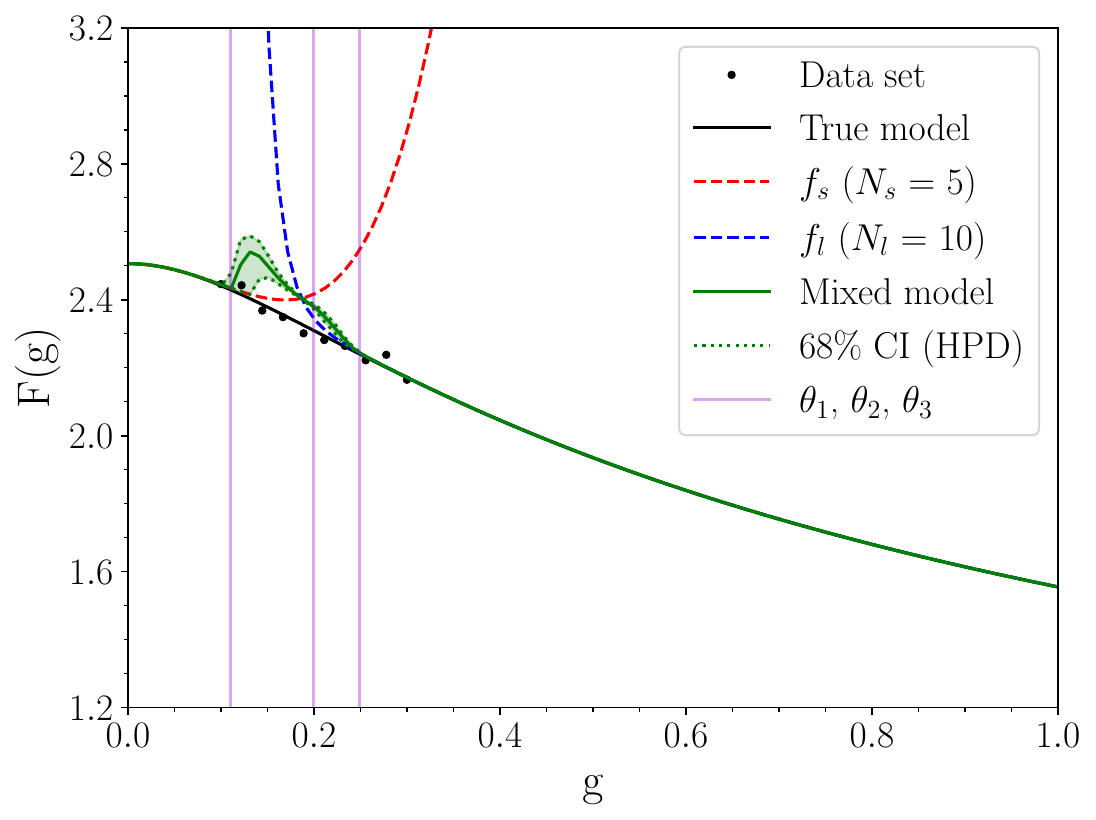}
	   \label{fig:sup_lmm_a}
	\end{minipage}}
 \hfill 	
  \subfloat[Truncation orders $N_{s}=5$ and $N_{l}=23$.]{
	\begin{minipage}[c][1\width]{
	   0.45\textwidth}
	   \centering
	   \includegraphics[width=1.1\textwidth]{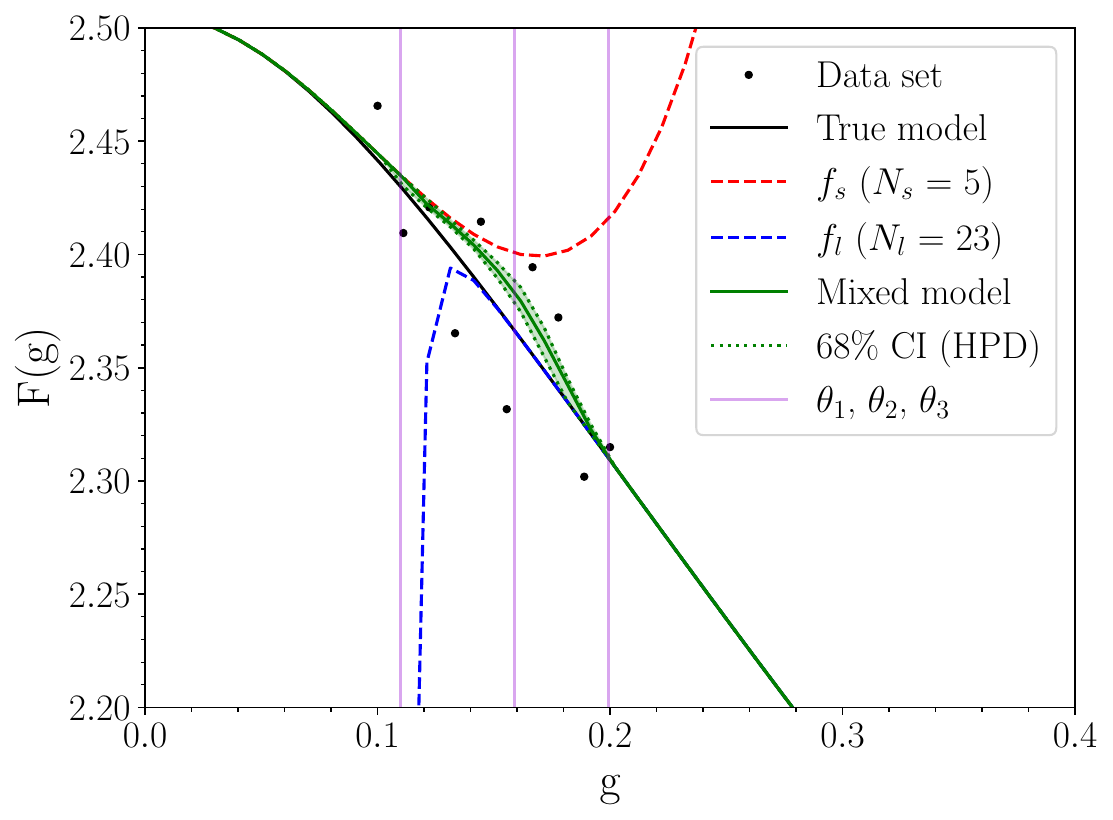}
	   \label{fig:sup_lmm_b}
	\end{minipage}}
    \caption{Supplemental truncation order plots calculated using the linear mixture model. As in Figs.~\ref{fig:pwclmm22}and~\ref{fig:pwclmm55}, the green solid line and shaded band represent the PPD (see Eq.~\eqref{eq:PPD_LMM}) and its associated 68\% credibility interval. The dashed red curve in each plot is $F_s^{N_s}(g)$ and the dashed blue curve is $F_l^{N_l}(g)$. The solid black curve represents the true model, with the black dots around it indicating the data sets generated for these cases of $N_s$ and $N_l$. The vertical purple lines represent the location in $g$ of each of the parameters of $\alpha(g,\bm{\theta})$.}
        \label{fig:supplemental_lmm}
\end{figure*}

We show two higher order plots from the linear mixture model BMM. As we increase $N_{s}$ and $N_{l}$, the gap between the two expansions decreases, leading to a shorter interpolation length between the two models. Results of these cases are presented in Fig.~\ref{fig:supplemental_lmm}. It is evident at first glance that the PPD shown there (in solid green), calculated using Eq.~\eqref{eq:PPD_LMM}, is not a good representation of the true curve in either Fig.~\ref{fig:sup_lmm_a} or \ref{fig:sup_lmm_b}. In both cases, the PPD 68\% credibility interval does not encompass 68\% of the true curve.

\subsection{Localized Bivariate Bayesian Model Mixing}

\begin{figure*}[htbp]
  \subfloat[Truncation orders $N_s=4$ and $N_l=4$.]{
	\begin{minipage}[c][1\width]{
	   0.45\textwidth}
	   \centering
	   \includegraphics[width=1.1\textwidth]{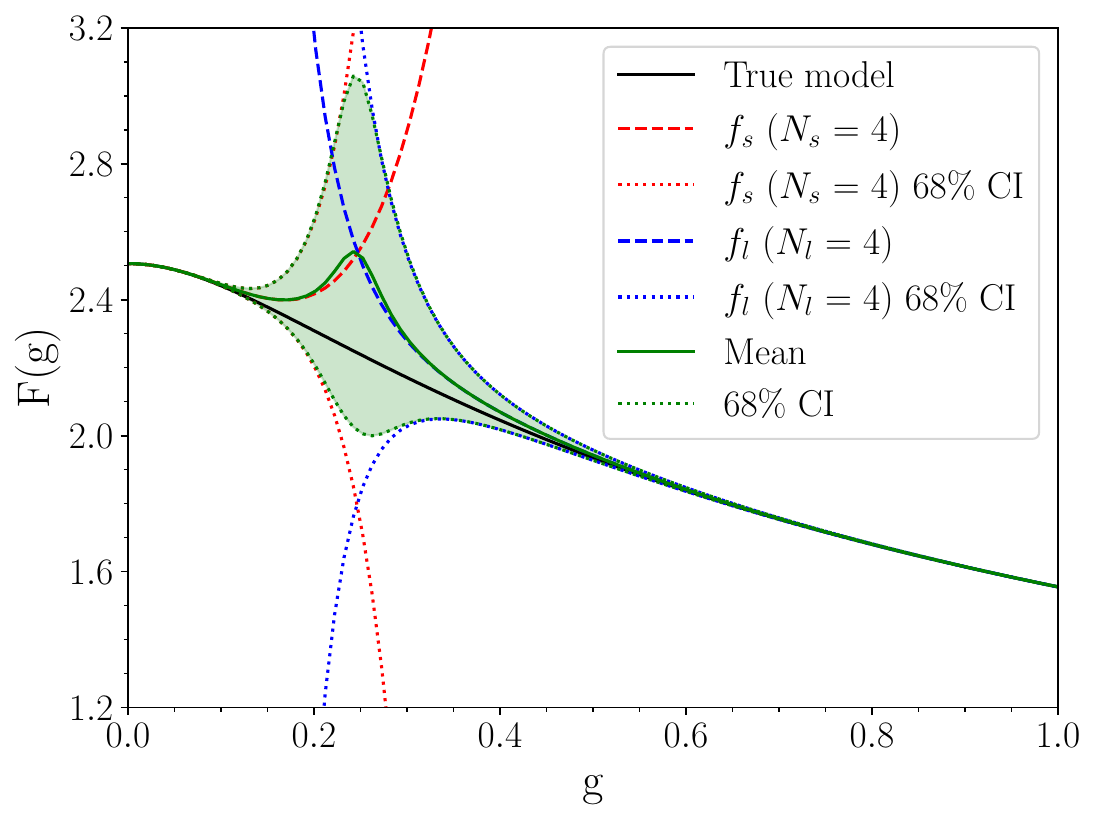}
	   \label{fig:sup_biv_1_a}
	\end{minipage}}
 \hfill 	
  \subfloat[Truncation orders $N_s=16$ and $N_l=16$.]{
	\begin{minipage}[c][1\width]{
	   0.45\textwidth}
	   \centering
	   \includegraphics[width=1.1\textwidth]{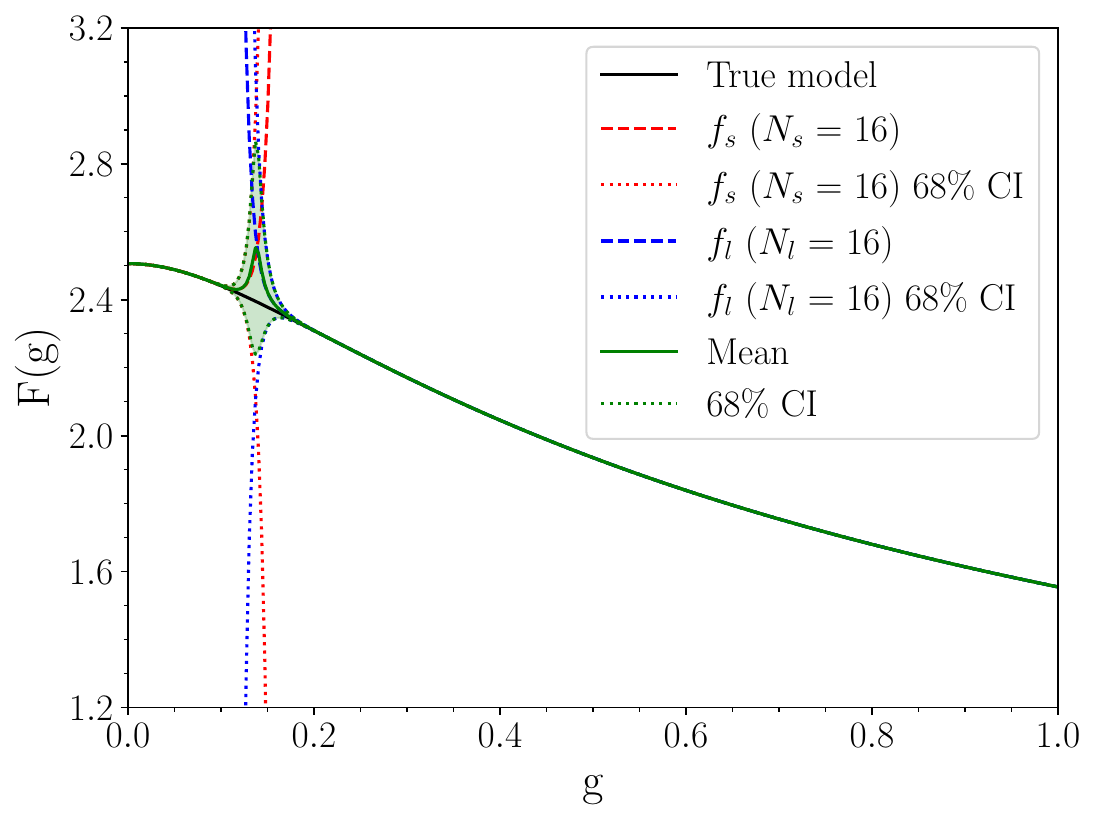}
	   \label{fig:sup_biv_1_b}
	\end{minipage}}
    \caption{Supplemental truncation order plots calculated in the bivariate Bayesian model mixing method (see Eq.~\eqref{eq:fdagger}). Here the green curve and light green band represent the PPD and 68\% credibility interval for the mixed model. The dashed red curve and dashed blue curve are the results for $F_s^{N_s}(g)$ and $F_l^{N_l}(g)$, respectively, with the dotted curves indicating their respective variances predicted by the next order in the series (see Sec.~\ref{sec:theoryerrormodel}).}
    \label{fig:supplemental_bivariate_1}
\end{figure*}

\begin{figure*}[htbp]
  \subfloat[Truncation orders $N_s=5$ and $N_l=10$.]{
	\begin{minipage}[c][1\width]{
	   0.45\textwidth}
	   \centering
	   \includegraphics[width=1.1\textwidth]{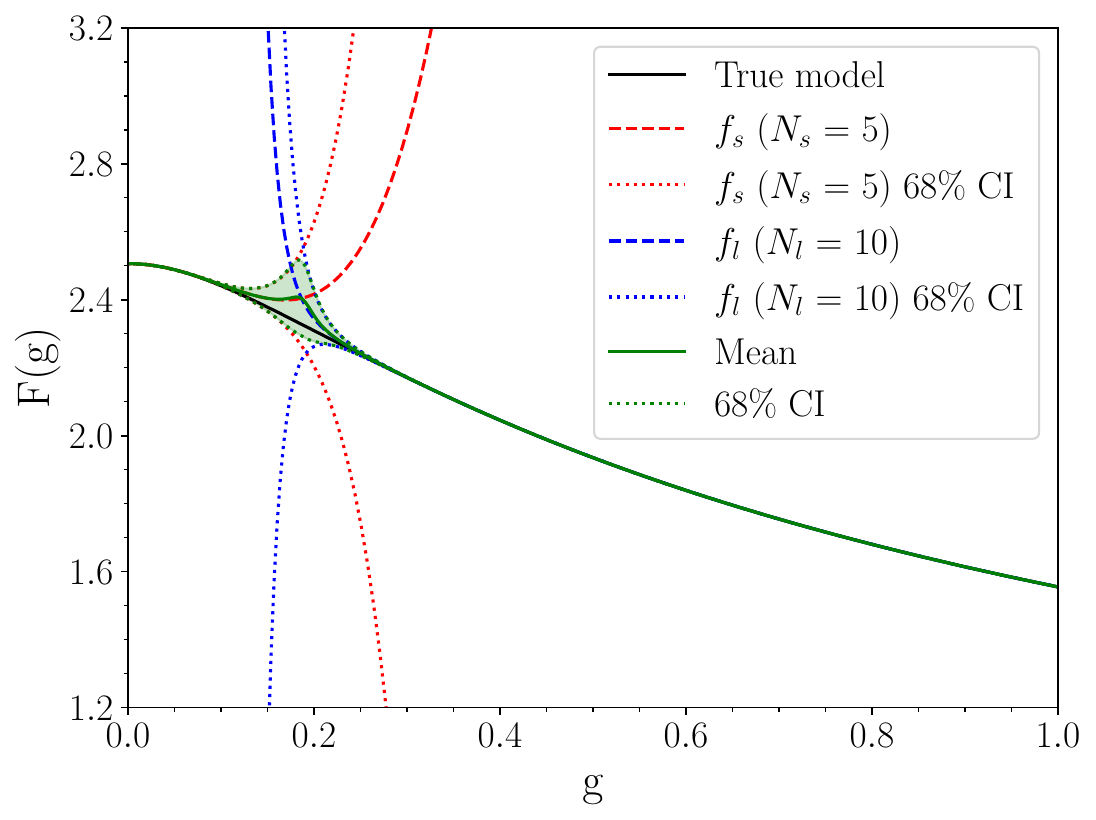}
	   \label{fig:sup_biv_2_a}
	\end{minipage}}
 \hfill 	
  \subfloat[Truncation orders $N_s=5$ and $N_l=23$.]{
	\begin{minipage}[c][1\width]{
	   0.45\textwidth}
	   \centering
	   \includegraphics[width=1.1\textwidth]{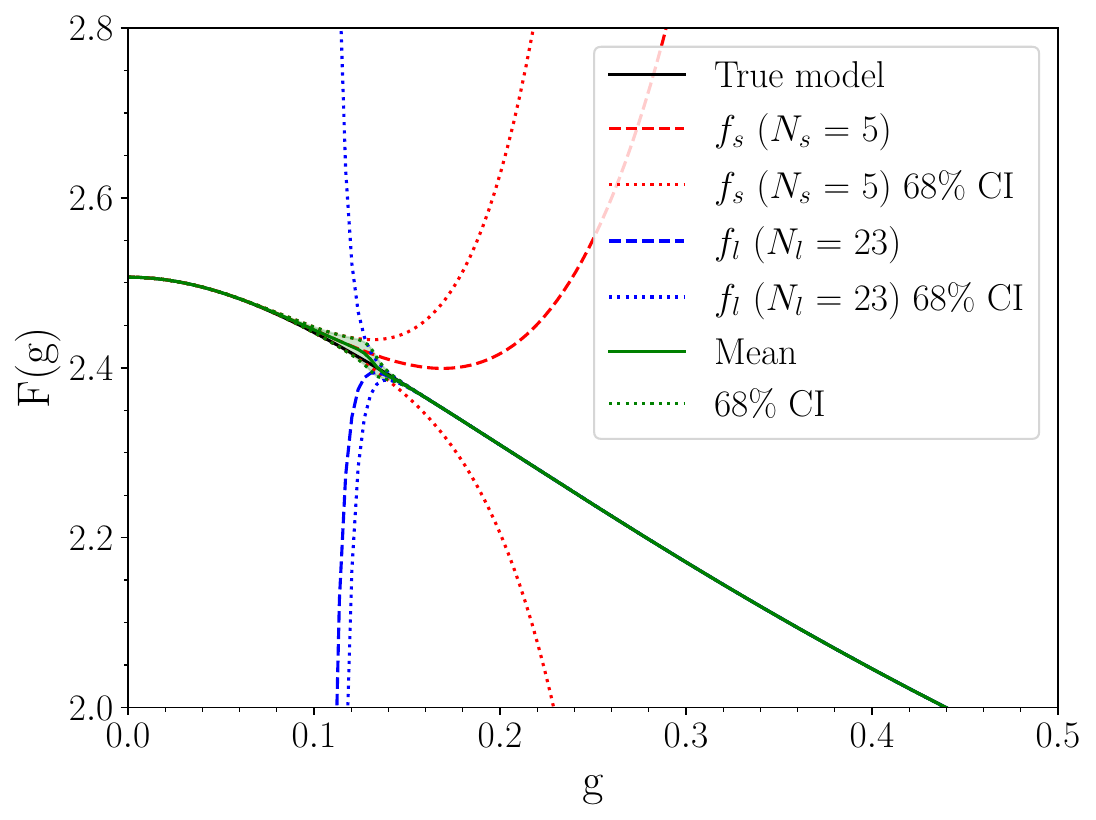}
	   \label{fig:sup_biv_2_b}
	\end{minipage}}
    \caption{Supplemental truncation order plots calculated in the bivariate Bayesian model mixing method, as in Fig.\ref{fig:supplemental_bivariate_1}.}
    \label{fig:supplemental_bivariate_2}
\end{figure*}

Several cases of different orders in $N_s$ and $N_l$ are presented in Figs.~\ref{fig:supplemental_bivariate_1} and~\ref{fig:supplemental_bivariate_2} as a supplement to those given in Sec.~\ref{sec:gaussians}. As the orders increase in both the small-$g$ and large-$g$ expansions, the gap gets smaller between the two, and with this, the bivariate mixed model PPD (shown in green) has less distance to travel to interpolate between the two models. If the orders of the expansions yield a case where one model is inflected upwards and the other downwards, the crossing via the PPD actually does rather well to map the true curve, as in Fig.~\ref{fig:sup_biv_2_b}. However, if both models are approaching $\infty$, the PPD traces out the curves where they cross quite precisely, as in Figs.~\ref{fig:sup_biv_1_a},~\ref{fig:sup_biv_1_b}, and~\ref{fig:sup_biv_2_a}. Nevertheless, this is not statistically wrong, as the 68\% credibility interval is quite conservative here and still covers the true curve in all three cases. 

\subsection{Localized multivariate Bayesian Model Mixing}

\begin{figure*}[p]
  \subfloat[Truncation orders $N_s=2$ and $N_l=4$.]{
	\begin{minipage}[c][1\width]{
	   0.45\textwidth}
	   \centering
	   \includegraphics[width=1.1\textwidth]{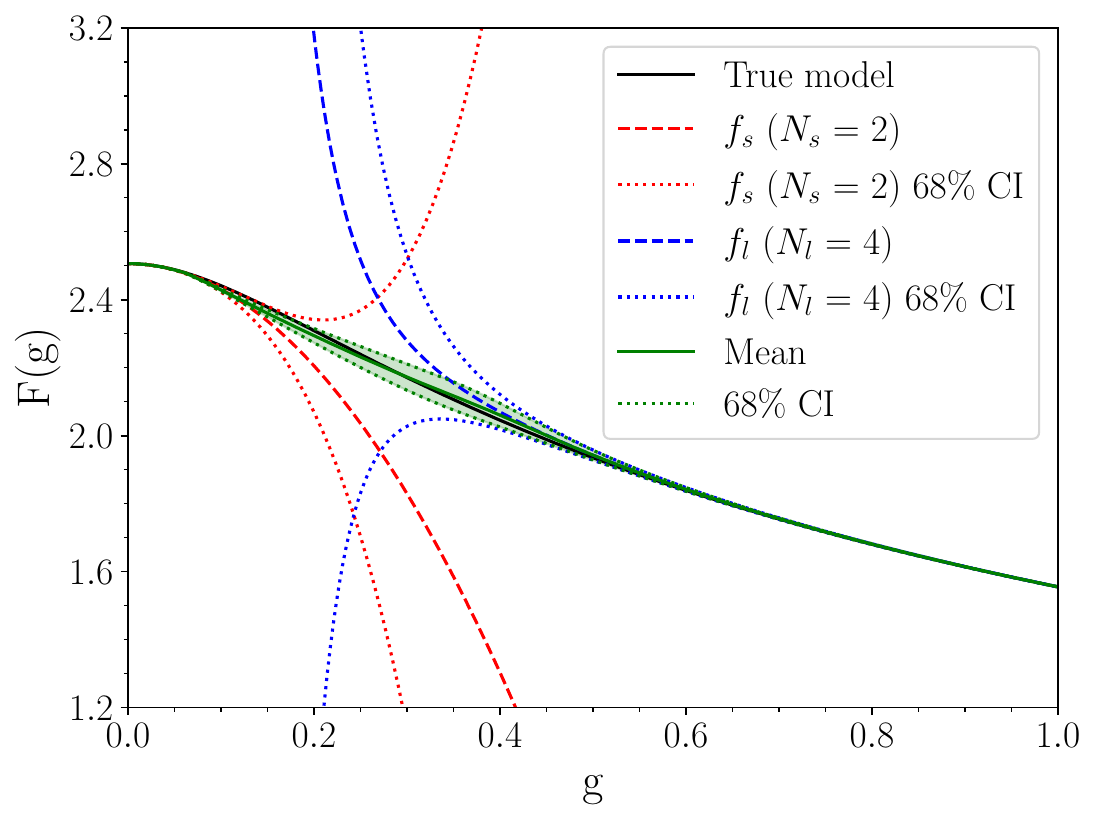}
	   \label{fig:sup_triv_1_a}
	\end{minipage}}
 \hfill 	
  \subfloat[Truncation orders $N_s=4$ and $N_l=4$.]{
	\begin{minipage}[c][1\width]{
	   0.45\textwidth}
	   \centering
	   \includegraphics[width=1.1\textwidth]{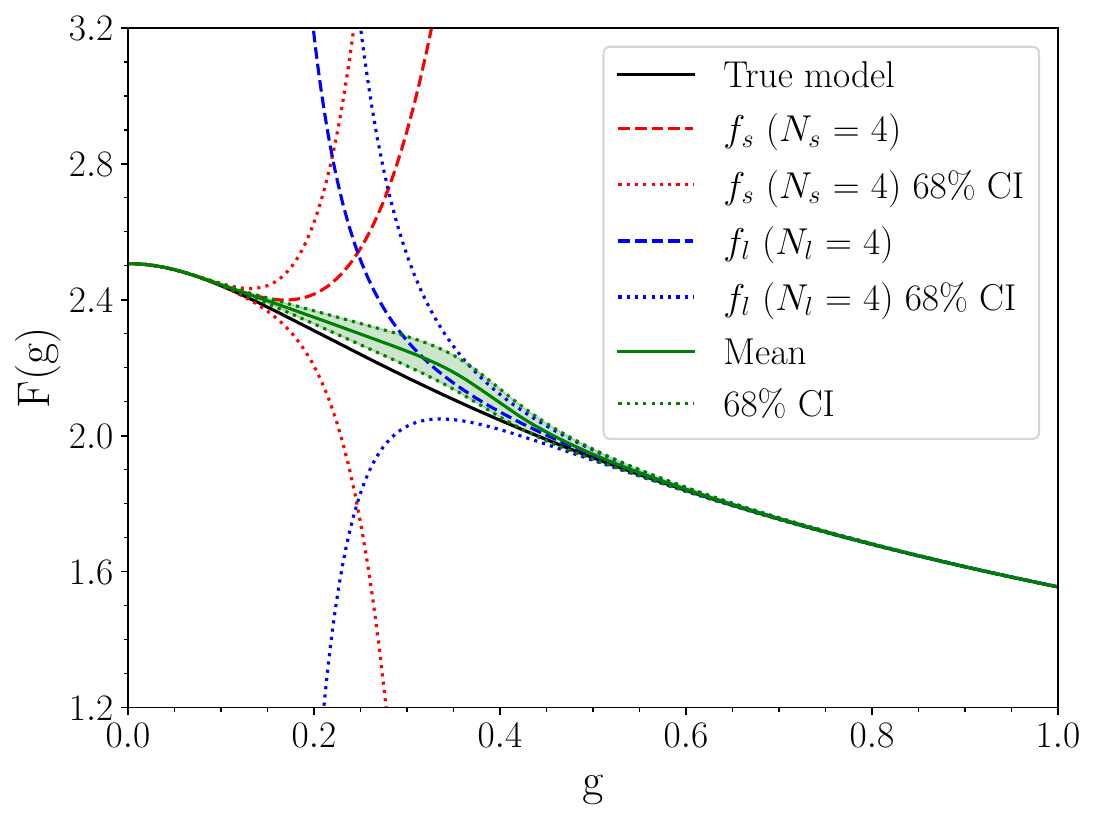}
	   \label{fig:sup_triv_1_b}
	\end{minipage}}
    \caption{Supplemental truncation order plots calculated in the trivariate Bayesian model mixing method (see Sec.~\ref{sec:gps}). Here the green curve and shaded green band represent the PPD of the mixed model and its associated 68\% credibility interval. The dashed red and blue curves are $F_{s}^{N_{s}}(g)$ and $F_{l}^{N_{l}}(g)$, respectively, with the dotted red and blue curves indicating their respective variances given by the next order in the series (see Sec.~\ref{sec:theoryerrormodel}). The solid black curve is the true model.}
    \label{fig:supplemental_trivariate_1}
\end{figure*}

\begin{figure*}[p]
  \subfloat[Truncation orders $N_s=5$ and $N_l=10$.]{
	\begin{minipage}[c][1\width]{
	   0.45\textwidth}
	   \centering
	   \includegraphics[width=1.1\textwidth]{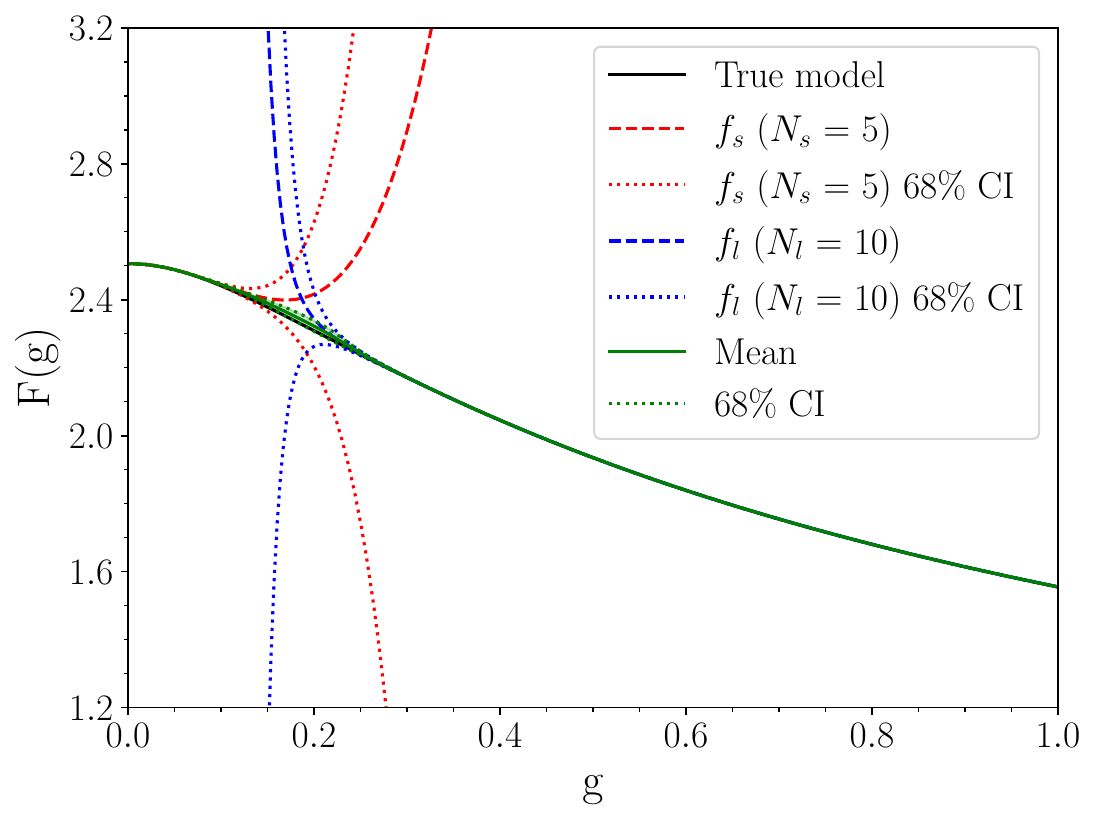}
	   \label{fig:sup_triv_2_a}
	\end{minipage}}
 \hfill 	
  \subfloat[Truncation orders $N_s=5$ and $N_l=23$.]{
	\begin{minipage}[c][1\width]{
	   0.45\textwidth}
	   \centering
	   \includegraphics[width=1.1\textwidth]{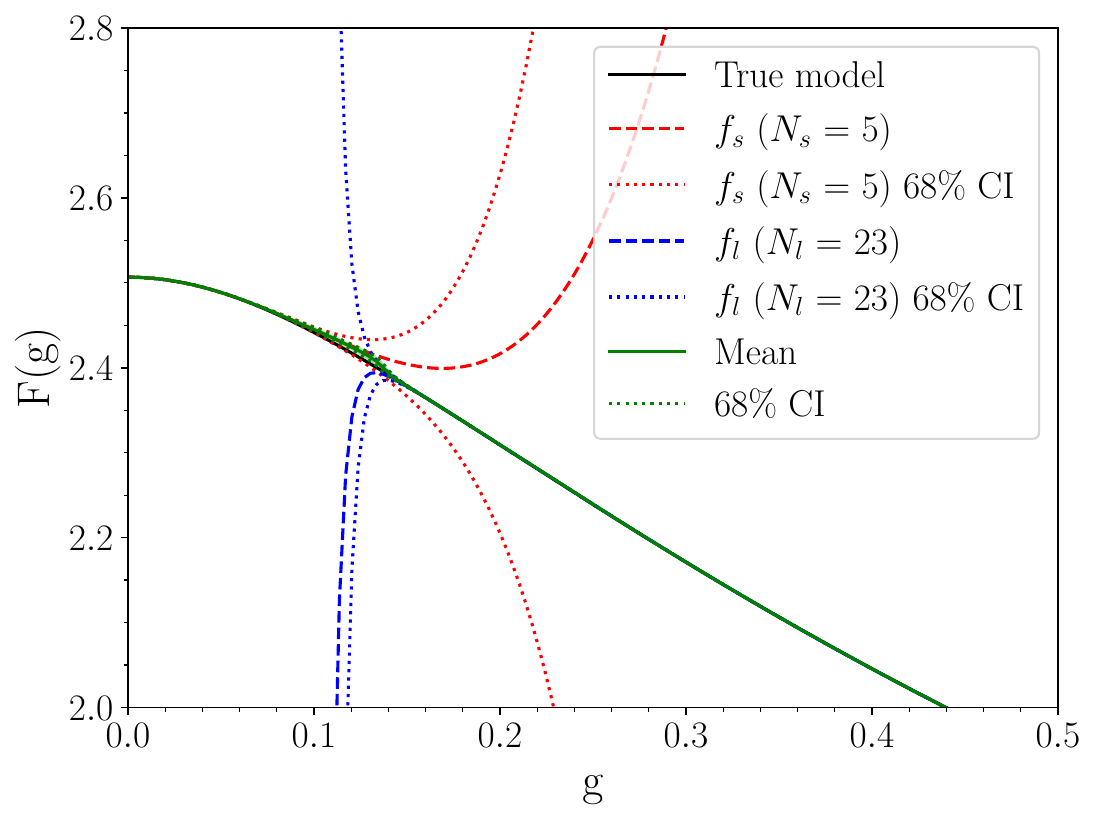}
	   \label{fig:sup_triv_2_b}
	\end{minipage}}
    \caption{Supplemental truncation order plots calculated in the trivariate Bayesian model mixing method, as in Fig.~\ref{fig:supplemental_trivariate_1}.}
    \label{fig:supplemental_trivariate_2}
\end{figure*}

\begin{figure*}[p]
  \subfloat[Truncation orders $N_s=8$ and $N_l=7$.]{
	\begin{minipage}[c][1\width]{
	   0.45\textwidth}
	   \centering
	   \includegraphics[width=1.1\textwidth]{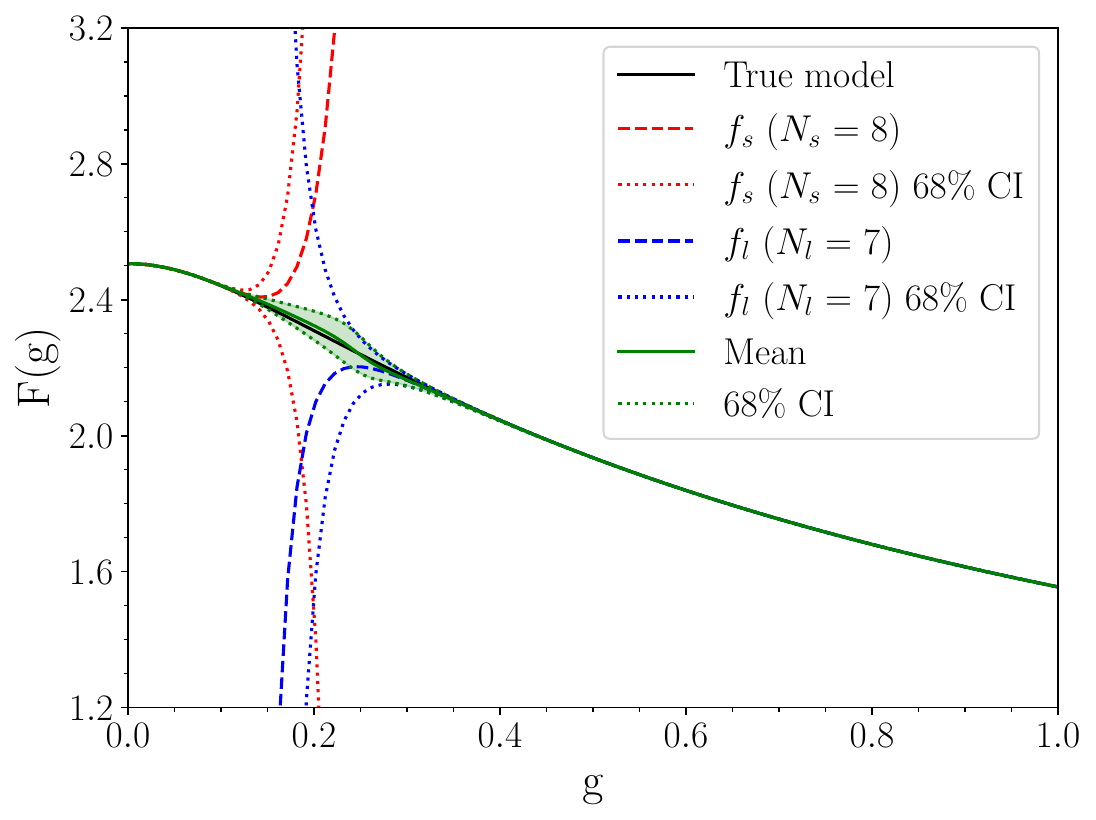}
	   \label{fig:sup_triv_3_a}
	\end{minipage}}
 \hfill 	
  \subfloat[Truncation orders $N_s=16$ and $N_l=16$.]{
	\begin{minipage}[c][1\width]{
	   0.45\textwidth}
	   \centering
	   \includegraphics[width=1.1\textwidth]{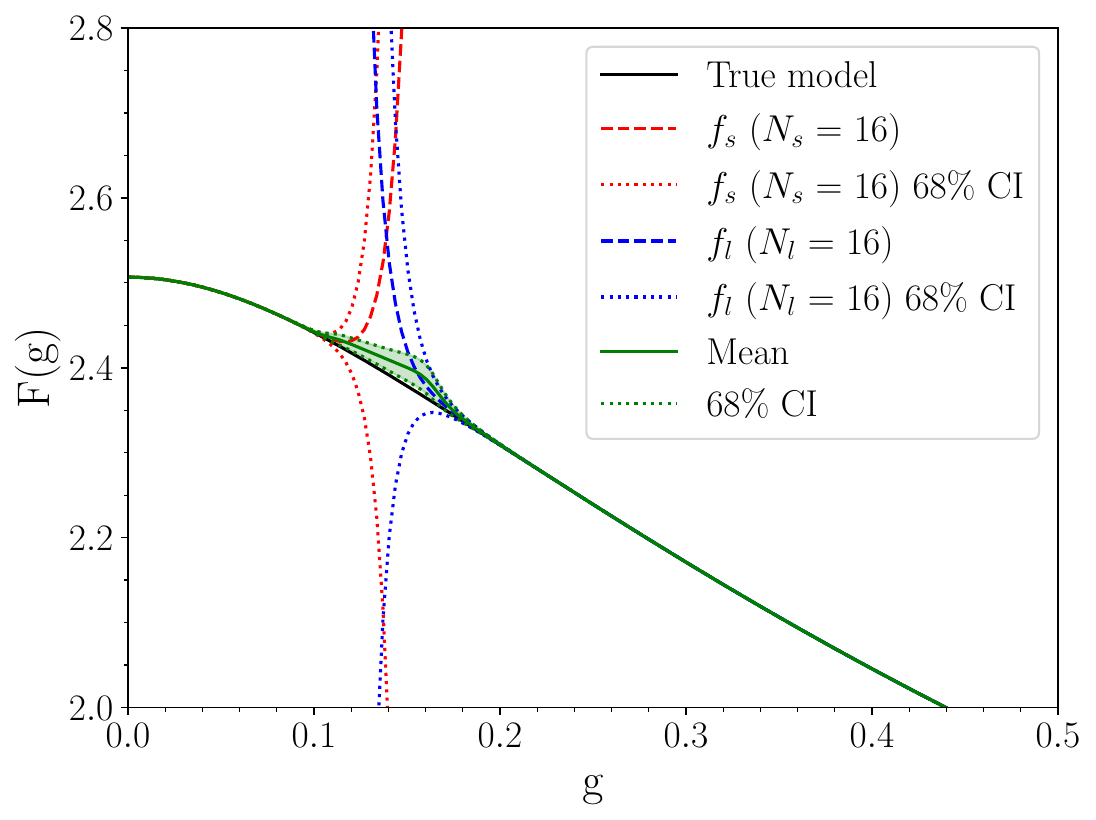}
	   \label{fig:sup_triv_3_b}
	\end{minipage}}
    \caption{Supplemental truncation order plots calculated in the trivariate Bayesian model mixing method, as in Figs.~\ref{fig:supplemental_trivariate_1} and~\ref{fig:supplemental_trivariate_3}.}
    \label{fig:supplemental_trivariate_3}
\end{figure*}

Supplemental cases of $N_s$ and $N_l$ are given here, to add to those in Sec.~\ref{sec:gps}. Figs.~\ref{fig:sup_triv_2_b} and~\ref{fig:sup_triv_3_b} represent higher truncation order cases, while Figs.~\ref{fig:supplemental_trivariate_1},~\ref{fig:sup_triv_2_a}, and~\ref{fig:sup_triv_3_a} give the PPD results (see Eq.~\eqref{eq:fdagger}) for trivariate model mixing that are highlighted in Table~\ref{tab:training_methods}, but not shown pictorially in the main text. All of the figures presented here use training method 2, with 4 total training points selected (see Sec.~\ref{sec:multiformalism} for a discourse on training the GP in this mixing method). 

\clearpage
\bibliography{bayesian_refs}

\begin{thebibliography}{32}%
\makeatletter
\providecommand \@ifxundefined [1]{%
 \@ifx{#1\undefined}
}%
\providecommand \@ifnum [1]{%
 \ifnum #1\expandafter \@firstoftwo
 \else \expandafter \@secondoftwo
 \fi
}%
\providecommand \@ifx [1]{%
 \ifx #1\expandafter \@firstoftwo
 \else \expandafter \@secondoftwo
 \fi
}%
\providecommand \natexlab [1]{#1}%
\providecommand \enquote  [1]{``#1''}%
\providecommand \bibnamefont  [1]{#1}%
\providecommand \bibfnamefont [1]{#1}%
\providecommand \citenamefont [1]{#1}%
\providecommand \href@noop [0]{\@secondoftwo}%
\providecommand \href [0]{\begingroup \@sanitize@url \@href}%
\providecommand \@href[1]{\@@startlink{#1}\@@href}%
\providecommand \@@href[1]{\endgroup#1\@@endlink}%
\providecommand \@sanitize@url [0]{\catcode `\\12\catcode `\$12\catcode
  `\&12\catcode `\#12\catcode `\^12\catcode `\_12\catcode `\%12\relax}%
\providecommand \@@startlink[1]{}%
\providecommand \@@endlink[0]{}%
\providecommand \url  [0]{\begingroup\@sanitize@url \@url }%
\providecommand \@url [1]{\endgroup\@href {#1}{\urlprefix }}%
\providecommand \urlprefix  [0]{URL }%
\providecommand \Eprint [0]{\href }%
\providecommand \doibase [0]{https://doi.org/}%
\providecommand \selectlanguage [0]{\@gobble}%
\providecommand \bibinfo  [0]{\@secondoftwo}%
\providecommand \bibfield  [0]{\@secondoftwo}%
\providecommand \translation [1]{[#1]}%
\providecommand \BibitemOpen [0]{}%
\providecommand \bibitemStop [0]{}%
\providecommand \bibitemNoStop [0]{.\EOS\space}%
\providecommand \EOS [0]{\spacefactor3000\relax}%
\providecommand \BibitemShut  [1]{\csname bibitem#1\endcsname}%
\let\auto@bib@innerbib\@empty
\bibitem [{\citenamefont {Glendenning}(1992)}]{Glendenning:1992vb}%
  \BibitemOpen
  \bibfield  {author} {\bibinfo {author} {\bibfnamefont {N.~K.}\ \bibnamefont
  {Glendenning}},\ }\bibfield  {title} {\bibinfo {title} {{First order phase
  transitions with more than one conserved charge: Consequences for neutron
  stars}},\ }\href {https://doi.org/10.1103/PhysRevD.46.1274} {\bibfield
  {journal} {\bibinfo  {journal} {Phys. Rev. D}\ }\textbf {\bibinfo {volume}
  {46}},\ \bibinfo {pages} {1274} (\bibinfo {year} {1992})}\BibitemShut
  {NoStop}%
\bibitem [{\citenamefont {Glendenning}(2001)}]{Glendenning:2001pe}%
  \BibitemOpen
  \bibfield  {author} {\bibinfo {author} {\bibfnamefont {N.~K.}\ \bibnamefont
  {Glendenning}},\ }\bibfield  {title} {\bibinfo {title} {{Phase transitions
  and crystalline structures in neutron star cores}},\ }\href
  {https://doi.org/10.1016/S0370-1573(00)00080-6} {\bibfield  {journal}
  {\bibinfo  {journal} {Phys. Rept.}\ }\textbf {\bibinfo {volume} {342}},\
  \bibinfo {pages} {393} (\bibinfo {year} {2001})}\BibitemShut {NoStop}%
\bibitem [{\citenamefont {McLerran}\ and\ \citenamefont
  {Reddy}(2019)}]{McLerran:2018hbz}%
  \BibitemOpen
  \bibfield  {author} {\bibinfo {author} {\bibfnamefont {L.}~\bibnamefont
  {McLerran}}\ and\ \bibinfo {author} {\bibfnamefont {S.}~\bibnamefont
  {Reddy}},\ }\bibfield  {title} {\bibinfo {title} {{Quarkyonic Matter and
  Neutron Stars}},\ }\href {https://doi.org/10.1103/PhysRevLett.122.122701}
  {\bibfield  {journal} {\bibinfo  {journal} {Phys. Rev. Lett.}\ }\textbf
  {\bibinfo {volume} {122}},\ \bibinfo {pages} {122701} (\bibinfo {year}
  {2019})},\ \Eprint {https://arxiv.org/abs/1811.12503} {arXiv:1811.12503
  [nucl-th]} \BibitemShut {NoStop}%
\bibitem [{\citenamefont {Han}\ \emph {et~al.}(2019)\citenamefont {Han},
  \citenamefont {Mamun}, \citenamefont {Lalit}, \citenamefont {Constantinou},\
  and\ \citenamefont {Prakash}}]{Han:2019bub}%
  \BibitemOpen
  \bibfield  {author} {\bibinfo {author} {\bibfnamefont {S.}~\bibnamefont
  {Han}}, \bibinfo {author} {\bibfnamefont {M.~A.~A.}\ \bibnamefont {Mamun}},
  \bibinfo {author} {\bibfnamefont {S.}~\bibnamefont {Lalit}}, \bibinfo
  {author} {\bibfnamefont {C.}~\bibnamefont {Constantinou}},\ and\ \bibinfo
  {author} {\bibfnamefont {M.}~\bibnamefont {Prakash}},\ }\bibfield  {title}
  {\bibinfo {title} {{Treating quarks within neutron stars}},\ }\href
  {https://doi.org/10.1103/PhysRevD.100.103022} {\bibfield  {journal} {\bibinfo
   {journal} {Phys. Rev. D}\ }\textbf {\bibinfo {volume} {100}},\ \bibinfo
  {pages} {103022} (\bibinfo {year} {2019})},\ \Eprint
  {https://arxiv.org/abs/1906.04095} {arXiv:1906.04095 [astro-ph.HE]}
  \BibitemShut {NoStop}%
\bibitem [{\citenamefont {Wellenhofer}\ \emph
  {et~al.}(2020{\natexlab{a}})\citenamefont {Wellenhofer}, \citenamefont
  {Phillips},\ and\ \citenamefont {Schwenk}}]{Wellenhofer:2020ykf}%
  \BibitemOpen
  \bibfield  {author} {\bibinfo {author} {\bibfnamefont {C.}~\bibnamefont
  {Wellenhofer}}, \bibinfo {author} {\bibfnamefont {D.~R.}\ \bibnamefont
  {Phillips}},\ and\ \bibinfo {author} {\bibfnamefont {A.}~\bibnamefont
  {Schwenk}},\ }\bibfield  {title} {\bibinfo {title} {{Constrained
  extrapolation problem and order-dependent mappings}},\ }\href
  {https://doi.org/10.1002/pssb.202000554} {\bibfield  {journal} {\bibinfo
  {journal} {Physica Status Solidi~(b)}\ }\textbf {\bibinfo {volume} {258}},\
  \bibinfo {pages} {2000554} (\bibinfo {year} {2020}{\natexlab{a}})},\ \Eprint
  {https://arxiv.org/abs/2011.02105} {arXiv:2011.02105 [cond-mat.quant-gas]}
  \BibitemShut {NoStop}%
\bibitem [{\citenamefont {Johns}\ \emph {et~al.}(1996)\citenamefont {Johns},
  \citenamefont {Ellis},\ and\ \citenamefont {Lattimer}}]{Johns:1996ht}%
  \BibitemOpen
  \bibfield  {author} {\bibinfo {author} {\bibfnamefont {S.~M.}\ \bibnamefont
  {Johns}}, \bibinfo {author} {\bibfnamefont {P.~J.}\ \bibnamefont {Ellis}},\
  and\ \bibinfo {author} {\bibfnamefont {J.~M.}\ \bibnamefont {Lattimer}},\
  }\bibfield  {title} {\bibinfo {title} {{Numerical approximation to the
  thermodynamic integrals}},\ }\href {https://doi.org/10.1086/178212}
  {\bibfield  {journal} {\bibinfo  {journal} {Astrophys. J.}\ }\textbf
  {\bibinfo {volume} {473}},\ \bibinfo {pages} {1020} (\bibinfo {year}
  {1996})},\ \Eprint {https://arxiv.org/abs/nucl-th/9604004}
  {arXiv:nucl-th/9604004} \BibitemShut {NoStop}%
\bibitem [{\citenamefont {Zheng}\ \emph {et~al.}(2021)\citenamefont {Zheng}
  \emph {et~al.}}]{CLAS:2021apd}%
  \BibitemOpen
  \bibfield  {author} {\bibinfo {author} {\bibfnamefont {X.}~\bibnamefont
  {Zheng}} \emph {et~al.} (\bibinfo {collaboration} {CLAS}),\ }\bibfield
  {title} {\bibinfo {title} {{Measurement of the proton spin structure at long
  distances}},\ }\href {https://doi.org/10.1038/s41567-021-01198-z} {\bibfield
  {journal} {\bibinfo  {journal} {Nature Phys.}\ }\textbf {\bibinfo {volume}
  {17}},\ \bibinfo {pages} {736} (\bibinfo {year} {2021})},\ \Eprint
  {https://arxiv.org/abs/2102.02658} {arXiv:2102.02658 [nucl-ex]} \BibitemShut
  {NoStop}%
\bibitem [{\citenamefont {Honda}(2014)}]{Honda:2014bza}%
  \BibitemOpen
  \bibfield  {author} {\bibinfo {author} {\bibfnamefont {M.}~\bibnamefont
  {Honda}},\ }\bibfield  {title} {\bibinfo {title} {{On Perturbation theory
  improved by Strong coupling expansion}},\ }\href
  {https://doi.org/10.1007/JHEP12(2014)019} {\bibfield  {journal} {\bibinfo
  {journal} {JHEP}\ }\textbf {\bibinfo {volume} {12}},\ \bibinfo {pages}
  {019}},\ \Eprint {https://arxiv.org/abs/1408.2960} {arXiv:1408.2960 [hep-th]}
  \BibitemShut {NoStop}%
\bibitem [{\citenamefont {Wellenhofer}\ \emph
  {et~al.}(2020{\natexlab{b}})\citenamefont {Wellenhofer}, \citenamefont
  {Phillips},\ and\ \citenamefont {Schwenk}}]{Wellenhofer:2020ylh}%
  \BibitemOpen
  \bibfield  {author} {\bibinfo {author} {\bibfnamefont {C.}~\bibnamefont
  {Wellenhofer}}, \bibinfo {author} {\bibfnamefont {D.~R.}\ \bibnamefont
  {Phillips}},\ and\ \bibinfo {author} {\bibfnamefont {A.}~\bibnamefont
  {Schwenk}},\ }\bibfield  {title} {\bibinfo {title} {{From weak to strong:
  Constrained extrapolation of perturbation series with applications to dilute
  Fermi systems}},\ }\href {https://doi.org/10.1103/PhysRevResearch.2.043372}
  {\bibfield  {journal} {\bibinfo  {journal} {Phys. Rev. Res.}\ }\textbf
  {\bibinfo {volume} {2}},\ \bibinfo {pages} {043372} (\bibinfo {year}
  {2020}{\natexlab{b}})},\ \Eprint {https://arxiv.org/abs/2006.01429}
  {arXiv:2006.01429 [cond-mat.quant-gas]} \BibitemShut {NoStop}%
\bibitem [{\citenamefont {Chan}\ \emph {et~al.}(2022)\citenamefont {Chan},
  \citenamefont {DeBoer}, \citenamefont {Furnstahl}, \citenamefont {Liyanage},
  \citenamefont {Nunes}, \citenamefont {Odell}, \citenamefont {Phillips},
  \citenamefont {Plumlee}, \citenamefont {Semposki}, \citenamefont {S\"urer},\
  and\ \citenamefont {Wild}}]{bandframework}%
  \BibitemOpen
  \bibfield  {author} {\bibinfo {author} {\bibfnamefont {M.~Y.-H.}\
  \bibnamefont {Chan}}, \bibinfo {author} {\bibfnamefont {R.~J.}\ \bibnamefont
  {DeBoer}}, \bibinfo {author} {\bibfnamefont {R.~J.}\ \bibnamefont
  {Furnstahl}}, \bibinfo {author} {\bibfnamefont {D.}~\bibnamefont {Liyanage}},
  \bibinfo {author} {\bibfnamefont {F.~M.}\ \bibnamefont {Nunes}}, \bibinfo
  {author} {\bibfnamefont {D.}~\bibnamefont {Odell}}, \bibinfo {author}
  {\bibfnamefont {D.~R.}\ \bibnamefont {Phillips}}, \bibinfo {author}
  {\bibfnamefont {M.}~\bibnamefont {Plumlee}}, \bibinfo {author} {\bibfnamefont
  {A.~C.}\ \bibnamefont {Semposki}}, \bibinfo {author} {\bibfnamefont
  {O.}~\bibnamefont {S\"urer}},\ and\ \bibinfo {author} {\bibfnamefont {S.~M.}\
  \bibnamefont {Wild}},\ }\href
  {https://github.com/bandframework/bandframework} {\emph {\bibinfo {title}
  {{BANDFramework: An} Open-Source Framework for {B}ayesian Analysis of Nuclear
  Dynamics}}},\ \bibinfo {type} {Tech. Rep.}\ \bibinfo {number} {Version
  0.2.0}\ (\bibinfo {year} {2022})\BibitemShut {NoStop}%
\bibitem [{\citenamefont {Plumlee}\ \emph {et~al.}(2021)\citenamefont
  {Plumlee}, \citenamefont {Özge Sürer},\ and\ \citenamefont
  {Wild}}]{surmise2021}%
  \BibitemOpen
  \bibfield  {author} {\bibinfo {author} {\bibfnamefont {M.}~\bibnamefont
  {Plumlee}}, \bibinfo {author} {\bibnamefont {Özge Sürer}},\ and\ \bibinfo
  {author} {\bibfnamefont {S.~M.}\ \bibnamefont {Wild}},\ }\href
  {https://surmise.readthedocs.io} {\emph {\bibinfo {title} {Surmise Users
  Manual}}},\ \bibinfo {type} {Tech. Rep.}\ \bibinfo {number} {Version 0.1.0}\
  (\bibinfo  {institution} {NAISE},\ \bibinfo {year} {2021})\BibitemShut
  {NoStop}%
\bibitem [{\citenamefont {Hoeting}\ \emph {et~al.}(1999)\citenamefont
  {Hoeting}, \citenamefont {Madigan}, \citenamefont {Raftery},\ and\
  \citenamefont {Volinsky}}]{Hoeting:1999abc}%
  \BibitemOpen
  \bibfield  {author} {\bibinfo {author} {\bibfnamefont {J.~A.}\ \bibnamefont
  {Hoeting}}, \bibinfo {author} {\bibfnamefont {D.}~\bibnamefont {Madigan}},
  \bibinfo {author} {\bibfnamefont {A.~E.}\ \bibnamefont {Raftery}},\ and\
  \bibinfo {author} {\bibfnamefont {C.~T.}\ \bibnamefont {Volinsky}},\
  }\bibfield  {title} {\bibinfo {title} {Bayesian model averaging: A
  tutorial},\ }\href {http://www.jstor.org/stable/2676803} {\bibfield
  {journal} {\bibinfo  {journal} {Statistical Science}\ }\textbf {\bibinfo
  {volume} {14}},\ \bibinfo {pages} {382} (\bibinfo {year} {1999})}\BibitemShut
  {NoStop}%
\bibitem [{\citenamefont {Yao}\ \emph {et~al.}(2018)\citenamefont {Yao},
  \citenamefont {Vehtari}, \citenamefont {Simpson},\ and\ \citenamefont
  {Gelman}}]{Yao:2018abc}%
  \BibitemOpen
  \bibfield  {author} {\bibinfo {author} {\bibfnamefont {Y.}~\bibnamefont
  {Yao}}, \bibinfo {author} {\bibfnamefont {A.}~\bibnamefont {Vehtari}},
  \bibinfo {author} {\bibfnamefont {D.}~\bibnamefont {Simpson}},\ and\ \bibinfo
  {author} {\bibfnamefont {A.}~\bibnamefont {Gelman}},\ }\bibfield  {title}
  {\bibinfo {title} {{Using Stacking to Average Bayesian Predictive
  Distributions (with Discussion)}},\ }\href
  {https://doi.org/10.1214/17-BA1091} {\bibfield  {journal} {\bibinfo
  {journal} {Bayesian Analysis}\ }\textbf {\bibinfo {volume} {13}},\ \bibinfo
  {pages} {917 } (\bibinfo {year} {2018})}\BibitemShut {NoStop}%
\bibitem [{\citenamefont {Yao}\ \emph {et~al.}(2021)\citenamefont {Yao},
  \citenamefont {Pirš}, \citenamefont {Vehtari},\ and\ \citenamefont
  {Gelman}}]{Yao:2021abc}%
  \BibitemOpen
  \bibfield  {author} {\bibinfo {author} {\bibfnamefont {Y.}~\bibnamefont
  {Yao}}, \bibinfo {author} {\bibfnamefont {G.}~\bibnamefont {Pirš}}, \bibinfo
  {author} {\bibfnamefont {A.}~\bibnamefont {Vehtari}},\ and\ \bibinfo {author}
  {\bibfnamefont {A.}~\bibnamefont {Gelman}},\ }\bibfield  {title} {\bibinfo
  {title} {{Bayesian Hierarchical Stacking: Some Models Are (Somewhere)
  Useful}},\ }\href {https://doi.org/10.1214/21-BA1287} {\bibfield  {journal}
  {\bibinfo  {journal} {Bayesian Analysis}\ ,\ \bibinfo {pages} {1 }} (\bibinfo
  {year} {2021})}\BibitemShut {NoStop}%
\bibitem [{\citenamefont {Fernández}\ and\ \citenamefont
  {Green}(2002)}]{BMM_local_10.2307/3088815}%
  \BibitemOpen
  \bibfield  {author} {\bibinfo {author} {\bibfnamefont {C.}~\bibnamefont
  {Fernández}}\ and\ \bibinfo {author} {\bibfnamefont {P.~J.}\ \bibnamefont
  {Green}},\ }\bibfield  {title} {\bibinfo {title} {Modelling spatially
  correlated data via mixtures: A bayesian approach},\ }\href
  {http://www.jstor.org/stable/3088815} {\bibfield  {journal} {\bibinfo
  {journal} {Journal of the Royal Statistical Society. Series B (Statistical
  Methodology)}\ }\textbf {\bibinfo {volume} {64}},\ \bibinfo {pages} {805}
  (\bibinfo {year} {2002})}\BibitemShut {NoStop}%
\bibitem [{\citenamefont {Phillips}\ \emph {et~al.}(2021)\citenamefont
  {Phillips}, \citenamefont {Furnstahl}, \citenamefont {Heinz}, \citenamefont
  {Maiti}, \citenamefont {Nazarewicz}, \citenamefont {Nunes}, \citenamefont
  {Plumlee}, \citenamefont {Pratola}, \citenamefont {Pratt}, \citenamefont
  {Viens},\ and\ \citenamefont {Wild}}]{Phillips:2020dmw}%
  \BibitemOpen
  \bibfield  {author} {\bibinfo {author} {\bibfnamefont {D.~R.}\ \bibnamefont
  {Phillips}}, \bibinfo {author} {\bibfnamefont {R.~J.}\ \bibnamefont
  {Furnstahl}}, \bibinfo {author} {\bibfnamefont {U.}~\bibnamefont {Heinz}},
  \bibinfo {author} {\bibfnamefont {T.}~\bibnamefont {Maiti}}, \bibinfo
  {author} {\bibfnamefont {W.}~\bibnamefont {Nazarewicz}}, \bibinfo {author}
  {\bibfnamefont {F.~M.}\ \bibnamefont {Nunes}}, \bibinfo {author}
  {\bibfnamefont {M.}~\bibnamefont {Plumlee}}, \bibinfo {author} {\bibfnamefont
  {M.~T.}\ \bibnamefont {Pratola}}, \bibinfo {author} {\bibfnamefont
  {S.}~\bibnamefont {Pratt}}, \bibinfo {author} {\bibfnamefont {F.~G.}\
  \bibnamefont {Viens}},\ and\ \bibinfo {author} {\bibfnamefont {S.~M.}\
  \bibnamefont {Wild}},\ }\bibfield  {title} {\bibinfo {title} {{Get on the
  BAND Wagon: A Bayesian Framework for Quantifying Model Uncertainties in
  Nuclear Dynamics}},\ }\href {https://doi.org/10.1088/1361-6471/abf1df}
  {\bibfield  {journal} {\bibinfo  {journal} {J. Phys. G}\ }\textbf {\bibinfo
  {volume} {48}},\ \bibinfo {pages} {072001} (\bibinfo {year} {2021})},\
  \Eprint {https://arxiv.org/abs/2012.07704} {arXiv:2012.07704 [nucl-th]}
  \BibitemShut {NoStop}%
\bibitem [{\citenamefont {Cacciari}\ and\ \citenamefont
  {Houdeau}(2011)}]{Cacciari:2011ze}%
  \BibitemOpen
  \bibfield  {author} {\bibinfo {author} {\bibfnamefont {M.}~\bibnamefont
  {Cacciari}}\ and\ \bibinfo {author} {\bibfnamefont {N.}~\bibnamefont
  {Houdeau}},\ }\bibfield  {title} {\bibinfo {title} {{Meaningful
  characterisation of perturbative theoretical uncertainties}},\ }\href
  {https://doi.org/10.1007/JHEP09(2011)039} {\bibfield  {journal} {\bibinfo
  {journal} {J. High Energy Phys}\ }\textbf {\bibinfo {volume} {09}},\ \bibinfo
  {pages} {039} (\bibinfo {year} {2011})},\ \Eprint
  {https://arxiv.org/abs/1105.5152} {arXiv:1105.5152} \BibitemShut {NoStop}%
\bibitem [{\citenamefont {Furnstahl}\ \emph {et~al.}(2015)\citenamefont
  {Furnstahl}, \citenamefont {Klco}, \citenamefont {Phillips},\ and\
  \citenamefont {Wesolowski}}]{Furnstahl:2015rha}%
  \BibitemOpen
  \bibfield  {author} {\bibinfo {author} {\bibfnamefont {R.~J.}\ \bibnamefont
  {Furnstahl}}, \bibinfo {author} {\bibfnamefont {N.}~\bibnamefont {Klco}},
  \bibinfo {author} {\bibfnamefont {D.~R.}\ \bibnamefont {Phillips}},\ and\
  \bibinfo {author} {\bibfnamefont {S.}~\bibnamefont {Wesolowski}},\ }\bibfield
   {title} {\bibinfo {title} {{Quantifying truncation errors in effective field
  theory}},\ }\href {https://doi.org/10.1103/PhysRevC.92.024005} {\bibfield
  {journal} {\bibinfo  {journal} {Phys. Rev. C}\ }\textbf {\bibinfo {volume}
  {92}},\ \bibinfo {pages} {024005} (\bibinfo {year} {2015})},\ \Eprint
  {https://arxiv.org/abs/1506.01343} {arXiv:1506.01343} \BibitemShut {NoStop}%
\bibitem [{\citenamefont {Melendez}\ \emph {et~al.}(2019)\citenamefont
  {Melendez}, \citenamefont {Furnstahl}, \citenamefont {Phillips},
  \citenamefont {Pratola},\ and\ \citenamefont
  {Wesolowski}}]{Melendez:2019izc}%
  \BibitemOpen
  \bibfield  {author} {\bibinfo {author} {\bibfnamefont {J.~A.}\ \bibnamefont
  {Melendez}}, \bibinfo {author} {\bibfnamefont {R.~J.}\ \bibnamefont
  {Furnstahl}}, \bibinfo {author} {\bibfnamefont {D.~R.}\ \bibnamefont
  {Phillips}}, \bibinfo {author} {\bibfnamefont {M.~T.}\ \bibnamefont
  {Pratola}},\ and\ \bibinfo {author} {\bibfnamefont {S.}~\bibnamefont
  {Wesolowski}},\ }\bibfield  {title} {\bibinfo {title} {{Quantifying
  Correlated Truncation Errors in Effective Field Theory}},\ }\href
  {https://doi.org/10.1103/PhysRevC.100.044001} {\bibfield  {journal} {\bibinfo
   {journal} {Phys. Rev. C}\ }\textbf {\bibinfo {volume} {100}},\ \bibinfo
  {pages} {044001} (\bibinfo {year} {2019})},\ \Eprint
  {https://arxiv.org/abs/1904.10581} {arXiv:1904.10581} \BibitemShut {NoStop}%
\bibitem [{\citenamefont {Ellenberg}(2014)}]{Ellenberg:2014ma}%
  \BibitemOpen
  \bibfield  {author} {\bibinfo {author} {\bibfnamefont {J.}~\bibnamefont
  {Ellenberg}},\ }\href@noop {} {\emph {\bibinfo {title} {{How Not to Be Wrong:
  The Power of Mathematical Thinking}}}},\ \bibinfo {edition} {8th}\ ed.\
  (\bibinfo  {publisher} {Penguin Books},\ \bibinfo {year} {2014})\BibitemShut
  {NoStop}%
\bibitem [{\citenamefont {Coleman}(2019)}]{Coleman}%
  \BibitemOpen
  \bibfield  {author} {\bibinfo {author} {\bibfnamefont {J.~R.}\ \bibnamefont
  {Coleman}},\ }\emph {\bibinfo {title} {Topics in Bayesian Computer Model
  Emulation and Calibration, with Applications to High-Energy Particle
  Collisions}},\ \href@noop {} {Ph.D. thesis},\ \bibinfo  {school} {Duke
  University} (\bibinfo {year} {2019})\BibitemShut {NoStop}%
\bibitem [{\citenamefont {Foreman-Mackey}\ \emph {et~al.}(2013)\citenamefont
  {Foreman-Mackey}, \citenamefont {Hogg}, \citenamefont {Lang},\ and\
  \citenamefont {Goodman}}]{Foreman_Mackey:2013aa}%
  \BibitemOpen
  \bibfield  {author} {\bibinfo {author} {\bibfnamefont {D.}~\bibnamefont
  {Foreman-Mackey}}, \bibinfo {author} {\bibfnamefont {D.~W.}\ \bibnamefont
  {Hogg}}, \bibinfo {author} {\bibfnamefont {D.}~\bibnamefont {Lang}},\ and\
  \bibinfo {author} {\bibfnamefont {J.}~\bibnamefont {Goodman}},\ }\bibfield
  {title} {\bibinfo {title} {emcee: The mcmc hammer},\ }\href
  {http://www.jstor.org/stable/10.1086/670067} {\bibfield  {journal} {\bibinfo
  {journal} {Publications of the Astronomical Society of the Pacific}\ }\textbf
  {\bibinfo {volume} {125}},\ \bibinfo {pages} {pp. 306} (\bibinfo {year}
  {2013})}\BibitemShut {NoStop}%
\bibitem [{\citenamefont {Wittgenstein}(1922)}]{wittgenstein-1922}%
  \BibitemOpen
  \bibfield  {author} {\bibinfo {author} {\bibfnamefont {L.}~\bibnamefont
  {Wittgenstein}},\ }\bibfield  {title} {\bibinfo {title} {Tractatus
  logico-philosophicus},\ }\href
  {http://scholar.google.de/scholar.bib?q=info:1G2GoIkyCZIJ:scholar.google.com/&output=citation&hl=de&ct=citation&cd=0}
  {\bibfield  {journal} {\bibinfo  {journal} {London: Routledge, 1981}\ }
  (\bibinfo {year} {1922})},\ \bibinfo {note} {proposition 7}\BibitemShut
  {NoStop}%
\bibitem [{\citenamefont {Pedregosa}\ \emph {et~al.}(2011)\citenamefont
  {Pedregosa}, \citenamefont {Varoquaux}, \citenamefont {Gramfort},
  \citenamefont {Michel}, \citenamefont {Thirion}, \citenamefont {Grisel},
  \citenamefont {Blondel}, \citenamefont {Prettenhofer}, \citenamefont {Weiss},
  \citenamefont {Dubourg}, \citenamefont {Vanderplas}, \citenamefont {Passos},
  \citenamefont {Cournapeau}, \citenamefont {Brucher}, \citenamefont {Perrot},\
  and\ \citenamefont {Duchesnay}}]{scikit-learn}%
  \BibitemOpen
  \bibfield  {author} {\bibinfo {author} {\bibfnamefont {F.}~\bibnamefont
  {Pedregosa}}, \bibinfo {author} {\bibfnamefont {G.}~\bibnamefont
  {Varoquaux}}, \bibinfo {author} {\bibfnamefont {A.}~\bibnamefont {Gramfort}},
  \bibinfo {author} {\bibfnamefont {V.}~\bibnamefont {Michel}}, \bibinfo
  {author} {\bibfnamefont {B.}~\bibnamefont {Thirion}}, \bibinfo {author}
  {\bibfnamefont {O.}~\bibnamefont {Grisel}}, \bibinfo {author} {\bibfnamefont
  {M.}~\bibnamefont {Blondel}}, \bibinfo {author} {\bibfnamefont
  {P.}~\bibnamefont {Prettenhofer}}, \bibinfo {author} {\bibfnamefont
  {R.}~\bibnamefont {Weiss}}, \bibinfo {author} {\bibfnamefont
  {V.}~\bibnamefont {Dubourg}}, \bibinfo {author} {\bibfnamefont
  {J.}~\bibnamefont {Vanderplas}}, \bibinfo {author} {\bibfnamefont
  {A.}~\bibnamefont {Passos}}, \bibinfo {author} {\bibfnamefont
  {D.}~\bibnamefont {Cournapeau}}, \bibinfo {author} {\bibfnamefont
  {M.}~\bibnamefont {Brucher}}, \bibinfo {author} {\bibfnamefont
  {M.}~\bibnamefont {Perrot}},\ and\ \bibinfo {author} {\bibfnamefont
  {E.}~\bibnamefont {Duchesnay}},\ }\bibfield  {title} {\bibinfo {title}
  {Scikit-learn: Machine learning in {P}ython},\ }\href@noop {} {\bibfield
  {journal} {\bibinfo  {journal} {Journal of Machine Learning Research}\
  }\textbf {\bibinfo {volume} {12}},\ \bibinfo {pages} {2825} (\bibinfo {year}
  {2011})}\BibitemShut {NoStop}%
\bibitem [{\citenamefont {scikit-learn developers}(2022)}]{scikit-doc}%
  \BibitemOpen
  \bibfield  {author} {\bibinfo {author} {\bibnamefont {scikit-learn
  developers}},\ }\href@noop {} {\bibinfo {title} {1.7.5. kernels for gaussian
  processes}},\ \bibinfo {howpublished}
  {https://scikit-learn.org/stable/modules/gaussian\_process.html\#kernels-for-gaussian-processes}
  (\bibinfo {year} {2007-2022}),\ \bibinfo {note} {accessed:
  2022-05-19}\BibitemShut {NoStop}%
\bibitem [{\citenamefont {Bastos}\ and\ \citenamefont
  {O'Hagan}(2009)}]{BastosDiagnosticsGaussianProcess2009}%
  \BibitemOpen
  \bibfield  {author} {\bibinfo {author} {\bibfnamefont {L.~S.}\ \bibnamefont
  {Bastos}}\ and\ \bibinfo {author} {\bibfnamefont {A.}~\bibnamefont
  {O'Hagan}},\ }\bibfield  {title} {\bibinfo {title} {Diagnostics for
  {{Gaussian Process Emulators}}},\ }\href
  {https://doi.org/10.1198/TECH.2009.08019} {\bibfield  {journal} {\bibinfo
  {journal} {Technometrics}\ }\textbf {\bibinfo {volume} {51}},\ \bibinfo
  {pages} {425} (\bibinfo {year} {2009})}\BibitemShut {NoStop}%
\bibitem [{\citenamefont {Ababou}\ \emph {et~al.}(1994)\citenamefont {Ababou},
  \citenamefont {Bagtzoglou},\ and\ \citenamefont {Wood}}]{Ababou:1994}%
  \BibitemOpen
  \bibfield  {author} {\bibinfo {author} {\bibfnamefont {R.}~\bibnamefont
  {Ababou}}, \bibinfo {author} {\bibfnamefont {A.~C.}\ \bibnamefont
  {Bagtzoglou}},\ and\ \bibinfo {author} {\bibfnamefont {E.~F.}\ \bibnamefont
  {Wood}},\ }\bibfield  {title} {\bibinfo {title} {{On the condition number of
  covariance matrices in kriging, estimation, and simulation of random
  fields}},\ }\href {https://doi.org/10.1007/BF02065878} {\bibfield  {journal}
  {\bibinfo  {journal} {Math. Geo.}\ }\textbf {\bibinfo {volume} {26}},\
  \bibinfo {pages} {99} (\bibinfo {year} {1994})}\BibitemShut {NoStop}%
\bibitem [{\citenamefont {Huth}\ \emph {et~al.}(2021)\citenamefont {Huth},
  \citenamefont {Wellenhofer},\ and\ \citenamefont {Schwenk}}]{Huth:2020ozf}%
  \BibitemOpen
  \bibfield  {author} {\bibinfo {author} {\bibfnamefont {S.}~\bibnamefont
  {Huth}}, \bibinfo {author} {\bibfnamefont {C.}~\bibnamefont {Wellenhofer}},\
  and\ \bibinfo {author} {\bibfnamefont {A.}~\bibnamefont {Schwenk}},\
  }\bibfield  {title} {\bibinfo {title} {{New equations of state constrained by
  nuclear physics, observations, and QCD calculations of high-density nuclear
  matter}},\ }\href {https://doi.org/10.1103/PhysRevC.103.025803} {\bibfield
  {journal} {\bibinfo  {journal} {Phys. Rev. C}\ }\textbf {\bibinfo {volume}
  {103}},\ \bibinfo {pages} {025803} (\bibinfo {year} {2021})},\ \Eprint
  {https://arxiv.org/abs/2009.08885} {arXiv:2009.08885 [nucl-th]} \BibitemShut
  {NoStop}%
\bibitem [{\citenamefont {Kejzlar}\ \emph {et~al.}(2020)\citenamefont
  {Kejzlar}, \citenamefont {Neufcourt}, \citenamefont {Nazarewicz},\ and\
  \citenamefont {Reinhard}}]{Kejzlar:2020vla}%
  \BibitemOpen
  \bibfield  {author} {\bibinfo {author} {\bibfnamefont {V.}~\bibnamefont
  {Kejzlar}}, \bibinfo {author} {\bibfnamefont {L.}~\bibnamefont {Neufcourt}},
  \bibinfo {author} {\bibfnamefont {W.}~\bibnamefont {Nazarewicz}},\ and\
  \bibinfo {author} {\bibfnamefont {P.-G.}\ \bibnamefont {Reinhard}},\
  }\bibfield  {title} {\bibinfo {title} {{Statistical aspects of nuclear mass
  models}},\ }\href {https://doi.org/10.1088/1361-6471/ab907c} {\bibfield
  {journal} {\bibinfo  {journal} {J. Phys. G}\ }\textbf {\bibinfo {volume}
  {47}},\ \bibinfo {pages} {094001} (\bibinfo {year} {2020})},\ \Eprint
  {https://arxiv.org/abs/2002.04151} {arXiv:2002.04151 [nucl-th]} \BibitemShut
  {NoStop}%
\bibitem [{\citenamefont {Chipman}\ \emph {et~al.}(2010)\citenamefont
  {Chipman}, \citenamefont {George},\ and\ \citenamefont
  {McCulloch}}]{chipman2010bart}%
  \BibitemOpen
  \bibfield  {author} {\bibinfo {author} {\bibfnamefont {H.~A.}\ \bibnamefont
  {Chipman}}, \bibinfo {author} {\bibfnamefont {E.~I.}\ \bibnamefont
  {George}},\ and\ \bibinfo {author} {\bibfnamefont {R.~E.}\ \bibnamefont
  {McCulloch}},\ }\bibfield  {title} {\bibinfo {title} {Bart: Bayesian additive
  regression trees},\ }\href@noop {} {\bibfield  {journal} {\bibinfo  {journal}
  {The Annals of Applied Statistics}\ }\textbf {\bibinfo {volume} {4}},\
  \bibinfo {pages} {266} (\bibinfo {year} {2010})}\BibitemShut {NoStop}%
\bibitem [{\citenamefont {Hill}\ \emph {et~al.}(2020)\citenamefont {Hill},
  \citenamefont {Linero},\ and\ \citenamefont
  {Murray}}]{bart_doi:10.1146/annurev-statistics-031219-041110}%
  \BibitemOpen
  \bibfield  {author} {\bibinfo {author} {\bibfnamefont {J.}~\bibnamefont
  {Hill}}, \bibinfo {author} {\bibfnamefont {A.}~\bibnamefont {Linero}},\ and\
  \bibinfo {author} {\bibfnamefont {J.}~\bibnamefont {Murray}},\ }\bibfield
  {title} {\bibinfo {title} {Bayesian additive regression trees: A review and
  look forward},\ }\href
  {https://doi.org/10.1146/annurev-statistics-031219-041110} {\bibfield
  {journal} {\bibinfo  {journal} {Annual Review of Statistics and Its
  Application}\ }\textbf {\bibinfo {volume} {7}},\ \bibinfo {pages} {251}
  (\bibinfo {year} {2020})}\BibitemShut {NoStop}%
\bibitem [{\citenamefont {Neufcourt}\ \emph {et~al.}(2018)\citenamefont
  {Neufcourt}, \citenamefont {Cao}, \citenamefont {Nazarewicz},\ and\
  \citenamefont {Viens}}]{Neufcourt:2018syo}%
  \BibitemOpen
  \bibfield  {author} {\bibinfo {author} {\bibfnamefont {L.}~\bibnamefont
  {Neufcourt}}, \bibinfo {author} {\bibfnamefont {Y.}~\bibnamefont {Cao}},
  \bibinfo {author} {\bibfnamefont {W.}~\bibnamefont {Nazarewicz}},\ and\
  \bibinfo {author} {\bibfnamefont {F.}~\bibnamefont {Viens}},\ }\bibfield
  {title} {\bibinfo {title} {{Bayesian approach to model-based extrapolation of
  nuclear observables}},\ }\href {https://doi.org/10.1103/PhysRevC.98.034318}
  {\bibfield  {journal} {\bibinfo  {journal} {Phys. Rev. C}\ }\textbf {\bibinfo
  {volume} {98}},\ \bibinfo {pages} {034318} (\bibinfo {year} {2018})},\
  \Eprint {https://arxiv.org/abs/1806.00552} {arXiv:1806.00552 [nucl-th]}
  \BibitemShut {NoStop}%
\end{thebibliography}%


\end{document}